\documentclass[11pt,a4paper]{article}
\pdfoutput=1
\usepackage{jheppub}
\allowdisplaybreaks
\usepackage{color}
\usepackage{amsmath}
\usepackage{graphicx}
\usepackage{esint}
\usepackage{slashed}
\usepackage[english]{babel}
\usepackage{multirow}
\usepackage{ulem}
\definecolor{or}{rgb}{0.95,0.65,0}

\newcommand{\bea}{\begin{eqnarray}}
\newcommand{\eea}{\end{eqnarray}}

\begin{document}

\title{
Improving heavy Dirac neutrino prospects at future hadron colliders using machine learning
}

\author[a]{Jie Feng\footnote{Moved to Massachusetts Institute of Technology (MIT), Cambridge, Massachusetts 02139, USA.},}
\author[b]{Mingqiu Li,}
\author[b,c]{Qi-Shu Yan,}
\author[a,f]{Yu-Pan Zeng,}
\author[a]{Hong-Hao Zhang,}
\author[d]{Yongchao Zhang,}
\author[c,e]{Zhijie Zhao}

\affiliation[a]{School of Physics, Sun Yat-Sen University, Guangzhou 510275, China}

\affiliation[b]{School of Physics Sciences, University of Chinese Academy of Sciences, Beijing 100049, China}

\affiliation[c]{Center for Future High Energy Physics, Institute of High Energy Physics, Chinese Academy of Sciences, Beijing 100049, China}

\affiliation[d]{School of Physics, Southeast University, Nanjing 211189, China}

\affiliation[e]{Deutsches Elektronen-Synchrotron DESY, 22607 Hamburg, Germany}

\affiliation[f]{School of Electronics and Information Engineering, Guangdong Ocean University, Zhanjiang 524088, China.}

\emailAdd{fengjie5@mail2.sysu.edu.cn}
\emailAdd{limingqiu17@mails.ucas.ac.cn}
\emailAdd{yanqishu@ucas.ac.cn}
\emailAdd{zengyp8@mail2.sysu.edu.cn }
\emailAdd{zhh98@mail.sysu.edu.cn }
\emailAdd{zhangyongchao@seu.edu.cn}
\emailAdd{zhaozhijie@ihep.ac.cn}



\abstract{
In this work, by using the machine learning methods, we study the sensitivities of heavy pseudo-Dirac neutrino $N$ in the inverse seesaw at the high-energy hadron colliders. The production process for the signal is $pp \to \ell N \to 3 \ell + E_T^{\rm miss}$, while the dominant background is $p p \to W Z \to 3 \ell + E_T^{\rm miss}$. We use either the Multi-Layer Perceptron  or the Boosted Decision Tree with Gradient Boosting to analyse the kinematic observables and optimize the discrimination of background and signal events. It is found that the reconstructed $Z$ boson mass and heavy neutrino mass from the charged leptons and missing transverse energy play crucial roles in separating the signal from backgrounds. The prospects of heavy-light neutrino mixing $|V_{\ell N}|^2$ (with $\ell = e,\,\mu$) are estimated by using machine learning at the hadron colliders with $\sqrt{s}=14$ TeV, 27 TeV, and 100 TeV, and it is found that $|V_{\ell N}|^2$ can be improved up to ${\cal O} (10^{-6})$ for heavy neutrino mass $m_N = 100$ GeV and ${\cal O} (10^{-4})$ for $m_N = 1$ TeV. 
}


\maketitle

\section{Introduction}

Seesaw mechanisms provide natural explanations of the tiny neutrino masses. However, in the canonical type-I seesaw~\cite{Minkowski:1977sc, Mohapatra:1979ia, Yanagida:1979as, GellMann:1980vs, Glashow:1979nm}, the Dirac Yukawa couplings connecting the left- and right-handed neutrinos are generally very small $Y_D \lesssim 10^{-6}$ if the heavy right-handed neutrinos $N$ are at or below the TeV-scale. With some specific flavor structures such as the discrete symmetries (see reviews in e.g. Refs.~\cite{Altarelli:2010gt, Ishimori:2010au, King:2013eh, Petcov:2017ggy, Xing:2020ijf, Feruglio:2019ybq}), the Yukawa couplings could be much larger, or even of order one, while fitting the neutrino masses $m_\nu \sim 0.1$ eV~\cite{Zyla:2020zbs}. Then the heavy neutrinos can mix with the active neutrinos with a sizable value, and can be produced at the high-energy hadron~\cite{Datta:1993nm, Han:2006ip, Bray:2007ru, delAguila:2007qnc, Kersten:2007vk, delAguila:2008cj, Atre:2009rg, Chen:2011hc,  BhupalDev:2012zg, Das:2012ze, Dev:2013wba, Das:2014jxa, Deppisch:2015qwa, Gago:2015vma, Dev:2015kca,  Das:2015toa, Kang:2015uoc, Degrande:2016aje,  Das:2016hof, Mitra:2016kov, Anamiati:2016uxp, Antusch:2016ejd, Das:2017pvt,  Das:2017nvm,  Accomando:2017qcs, Das:2017gke, Cai:2017mow, Bhardwaj:2018lma, Helo:2018qej, Das:2018hph, Deppisch:2018eth, Pascoli:2018rsg, Cottin:2018nms,  Abada:2018sfh, Drewes:2019fou, Drewes:2021nqr}, lepton~\cite{Fargion:1995qb, delAguila:2005ssc, delAguila:2005pin,  Das:2012ze, Antusch:2015mia, Banerjee:2015gca, Asaka:2015oia, Antusch:2015gjw,  Antusch:2016ejd, Antusch:2016vyf, Cai:2017mow, Biswal:2017nfl, Antusch:2017pkq, Hernandez:2018cgc, Deppisch:2018eth, Chakraborty:2018khw, Han:2021pun, Drewes:2021nqr}, and lepton-hadron~\cite{Mondal:2015zba, Antusch:2016ejd,  Lindner:2016lxq, Mondal:2016kof, Cai:2017mow} colliders. As a result of the Majorana nature of the heavy neutrinos, the ``smoking-gun'' signal of heavy neutrinos at hadron colliders would be same-sign dilepton plus two jets $\ell^\pm \ell^\pm jj$, which is clearly a lepton number violating (LNV) signature beyond the standard model (SM) \cite{Pilaftsis:1991ug,Datta:1993nm,Almeida:2000pz,Han:2006ip,Bray:2007ru,Atre:2009rg}. 

In light of the small Yukawa couplings in the electroweak scale type-I seesaw, we consider the TeV-scale inverse seesaw, in which the small neutrino masses are governed by a small $\mu_S$ parameter in the heavy neutrino sector~\cite{Mohapatra:1986aw, Mohapatra:1986bd}. In the basic version of inverse seesaw, the heavy neutrino states are mass degenerate and form pseudo-Dirac fermion pairs, 
while the Majorana components are highly suppressed by the $\mu_S$ parameter. Different from the canonical type-I seesaw, the Yukawa couplings of heavy neutrinos with the active neutrinos can be of order one, thus mixing sizably with the active neutrinos and inducing very rich phenomenology. 
Recent studies of $N$ at the hadron colliders can be found e.g. in Refs.~\cite{delAguila:2008cj,  Chen:2011hc, Das:2012ze, BhupalDev:2012zg, Das:2014jxa,  Dev:2015kca, Deppisch:2015qwa, Das:2015toa, Antusch:2016ejd, Das:2016hof,  Anamiati:2016uxp,  Das:2017pvt,  Das:2018hph, Bhardwaj:2018lma, Drewes:2019byd}. For a relatively light $N$, say with mass $m_N \lesssim 1$ TeV, the dominate production channel at hadron colliders is through the charged-current Drell-Yan (DY) process $pp \to W^{\ast} \to \ell N$, with the subsequent decay $N \to \ell W$, assuming $N$ is heavier than the $W$ boson. Then the leptonic decays of $W$ boson in the final state will result in the trilepton signatures at the leading-order, i.e. three charged leptons plus significant missing transverse energy (MET). As the production of $N$ at high-energy colliders is induced by the heavy-light neutrino mixing, the corresponding mixing angle $|V_{\ell N}|^2$ can be directly measured, or constrained, in the trilepton signatures~\cite{Das:2014jxa, Das:2015toa,  Das:2016hof, Antusch:2016ejd, Das:2017pvt,  Das:2018hph, Bhardwaj:2018lma}. This is largely complementary to the limits on $|V_{\ell N}|^2$ from the high-intensity beam-dump experiments, meson decay data, beta decays, neutrinoless double-beta decays, reactor neutrino experiments, neutrino oscillation data, electroweak precision data (EWPD), direct high-energy collider searches in other channels e.g. the process $W^\pm W^\pm \to \ell^\pm \ell^\pm$, and the cosmological and astrophysical observations~\cite{Atre:2009rg, Deppisch:2015qwa, deGouvea:2015euy, Chrzaszcz:2019inj, Bolton:2019pcu, Fuks:2020att}. Furthermore, heavy neutrinos at the hundreds of GeV scale can be used to explain the muon $g-2$ discrepancy~\cite{Cirigliano:2021peb}.


In this paper, we apply machine learning (ML) to the searches of heavy pseudo-Dirac neutrinos in the inverse seesaw at the High-Luminosity LHC (HL-LHC) 14 TeV, High-Energy LHC (HE-LHC) 27 TeV and future 100 TeV colliders such as the Future Circular Collider (FCC-hh)~\cite{FCC-hh} and the Super Proton-Proton Collider (SPPC)~\cite{Tang:2015qga}. In particular, we use either the Multi-Layer Perceptron (MLP)~\cite{Fukushima:2013b} or the Boosted Decision Tree with Gradient boosting (BDTG) \cite{10.2307/2699986} technique in the Toolkit for Multivariate Data Analysis (TMVA) package \cite{2007physics...3039H} to define the multi-variate estimators to separate the signal from SM backgrounds. Combining some kinematic distributions of the charged leptons, MET and extra jets (see Figs.~\ref{fig:var_global} to \ref{fig:var_jet}), it is found that the discrimination power of these two methods are close (see Fig.~\ref{fig:estimator}). It is promising that the usage of ML can improve significantly the prospects of heavy-light neutrino mixing $|V_{\ell N}|^2$ at the high-energy hadron colliders (see Fig.~\ref{fig:Bln}). 
For instance, for the heavy neutrino mass $m_N = 200$ GeV, the mixing angles $|V_{eN}|^2$ and $|V_{\mu N}|^2$ can be probed, respectively, up to $9.3\times10^{-6}$ and $9.0\times10^{-6}$ at the future 100 TeV collider. This improves e.g. the sensitivities of $2.2\times10^{-4}$ and $2.1\times10^{-4}$ from Ref.~\cite{Pascoli:2018heg} with dynamic jet vetoes. Remarkably, for $m_N \lesssim 1$ TeV, the sensitivities at the 27 TeV and 100 TeV colliders are (much) better than the indirect upper limits $|V_{eN}|^2 < 2.2\times10^{-3}$ and $|V_{\mu N}|^2 < 9.0 \times10^{-4}$ from EWPD~\cite{delAguila:2008pw, Akhmedov:2013hec, Antusch:2014woa, Blennow:2016jkn}. 

The rest of the paper is organized as follows: the inverse seesaw is introduced in Section~\ref{sec:model}. The hadron collider analysis details are presented in Section~\ref{sec:analysis}, where the Monte Carlo (MC) signal and background generation is described in Section~\ref{sec:event}, all the distinguishing variables are defined in Section~\ref{sect:variables}, and the ML analysis is performed in Section~\ref{sect:ML}. The resultant sensitivities at the HL-LHC, HE-LHC and the future 100 TeV collider are obtained in Section~\ref{sec:sensitivity}, before we summarize and conclude in Section~\ref{sec:conclusion}. Some details for the longitudinal momentum of neutrino at the hadron colliders are given in the Appendix.



\section{Inverse seesaw}
\label{sec:model}


The inverse seesaw was proposed in the 1980s in Refs.~\cite{Mohapatra:1986aw, Mohapatra:1986bd}. In this model, two sets of fermions $N_{R,\alpha}$ and $S_{L,\beta}$ (with $\alpha$ and $\beta$ the flavor indices) are introduced, which are singlets under the SM and carry lepton numbers $L(N_R)=L(S_L) = 1$. For simplicity we consider the symmetric case with three pairs of $N_R$ and $S_L$. The Yukawa Lagrangian is given by
\begin{eqnarray}
-\mathcal{L}_{Y} \ = \ 
Y_{\alpha\beta} \bar{L}_\alpha \Phi N_{R,\,\beta} 
+ M_{N,\,\alpha\beta} \bar{S}_{L,\,\alpha} N_{R,\,\beta} + \frac{1}{2}\mu_{S,\,\alpha\beta}\bar{S}_{L,\,\alpha} S^C_{L,\,\beta}
+{\rm H.c.} \; ,
\label{eqn:Lagrangian}
\end{eqnarray}
where $L_\alpha = (\nu_\alpha,\; \ell_\alpha)^{\sf T}$ is the SM lepton doublet, $\Phi$ is the SM Higgs doublet, $S_L^C\equiv S_L^{\sf T}C^{-1}$ is the charge conjugate of $S_L$, $M_N$ is a Dirac mass term, and $\mu_S$ is the only Majorana mass term. After the electroweak symmetry breaking, the SM Higgs obtain its nonzero vacuum expectation value $\langle \Phi \rangle = (0, v/\sqrt2)^{\sf T}$, which leads to the Dirac mass term $M_D = Yv/\sqrt2$. In the flavor basis $\{\nu^C_{},\; N_{R},\; S^C_{L}\}$, the Lagrangian~(\ref{eqn:Lagrangian}) gives rise to the
full neutrino mass matrix 
\begin{eqnarray}
\label{mm}
{\cal M}_\nu \ = \ \begin{pmatrix}
{\mathbf 0} & M_D & {\mathbf 0} \\
M_D^{\sf T} & {\mathbf 0} & M_N^{\sf T} \\
{\mathbf 0} & M_N & \mu_S
\end{pmatrix}  \,.
\end{eqnarray}
In general, $M_D$, $M^{}_N$, and $\mu_S$ are all $3\times 3$ matrices, and the full mass matrix ${\cal M}_\nu$ in Eq.~(\ref{mm}) is a $9 \times 9$ matrix. 
In the limit of $||\mu_S|| \ll ||M_D|| \ll ||M_N||$ 
(with $||x|| \equiv \sqrt{{\rm tr} (x^\dagger x)}$ the positive norm of the matrix $x$), 
the mass matrix in Eq.~(\ref{mm}) can be block diagonalized, which leads to the light neutrino mass matrix
\begin{eqnarray}
M_\nu  \ = \ M_DM_N^{-1} \mu_S \: (M_N^{-1})^{\sf T} M_D^{\sf T} \,.
\label{eqn:mnu}
\end{eqnarray}
Different from the canonical type-I seesaw, the tiny neutrino masses are proportional to the small LNV parameter $\mu_S$ in the inverse seesaw. In light of the 't Hooft's naturalness criteria~\cite{tHooft:1979rat}, the LNV parameter $\mu_S$ is small, which provides a natural explanation for the small neutrino masses in the SM. The masses of heavy fermions $N_R$ and $S_L$ are determined predominantly by the matrix $M_N$. In particular, they form quasi-Dirac pairs $N_i$ and $\bar{N}_i$ ($i=1,\,2,\,3$) with mass eigenvalues at the order of $||M_N \mp \mu_S/2||$ with a small splitting of order $||\mu_S||$. 

In presence of the Majorana mass term $\mu_R \bar{N}_R N_R^C$ in the matrix (\ref{mm}), the active neutrino masses will obtain 1-loop contribution from the electroweak radiative corrections~\cite{Dev:2012sg, Dev:2012bd}. Furthermore, in the limit of $||\mu_S|| \ll ||M_D|| \ll ||M_N||  \ll ||\mu_R||$, the heavy mass eigenstates $N$ are nearly pure Majorana states with mass at the scale of $||\mu_R||$ as in the canonical type-I seesaw. In this case, as a result of the Majorana nature of $N$, we can expect equal number of lepton number conserving and LNV events from $N$ decay. For intermediate values of $\mu_R$, i.e. $||\mu_R|| \sim ||M_N||$, we can have scenarios with varying degree of LNV. In particular, the ratio of same-sign dilepton signal $pp \to \ell_\alpha^\pm N \ \to \ \ell_\alpha^\pm \ell_\beta^\pm W^\mp$ to the opposite-sign dilepton signal $\ell_\alpha^\pm \ell_\beta^\mp W^\pm$ can be anywhere between 0 and 1, depending on the ratio $||\mu_R||^2/||M_N||^2$. This will have important consequences for LNV signals at the high-energy colliders~\cite{Dev:2015pga}. For the same-sign and opposite-sign dilepton signatures, the SM backgrounds are different. For simplicity, we will consider here only the lepton number conserving signals from the heavy pseudo-Dirac neutrinos in the basic inverse seesaw scenario in Eq.~(\ref{mm}).

\begin{figure}[t]
  \centering
 \includegraphics[width=0.30\textwidth]{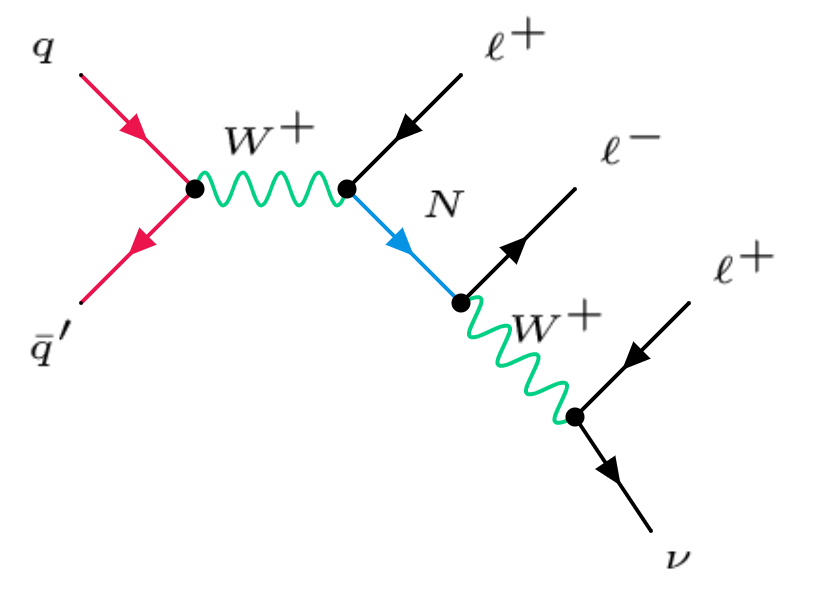}
  \includegraphics[width=0.34\textwidth]{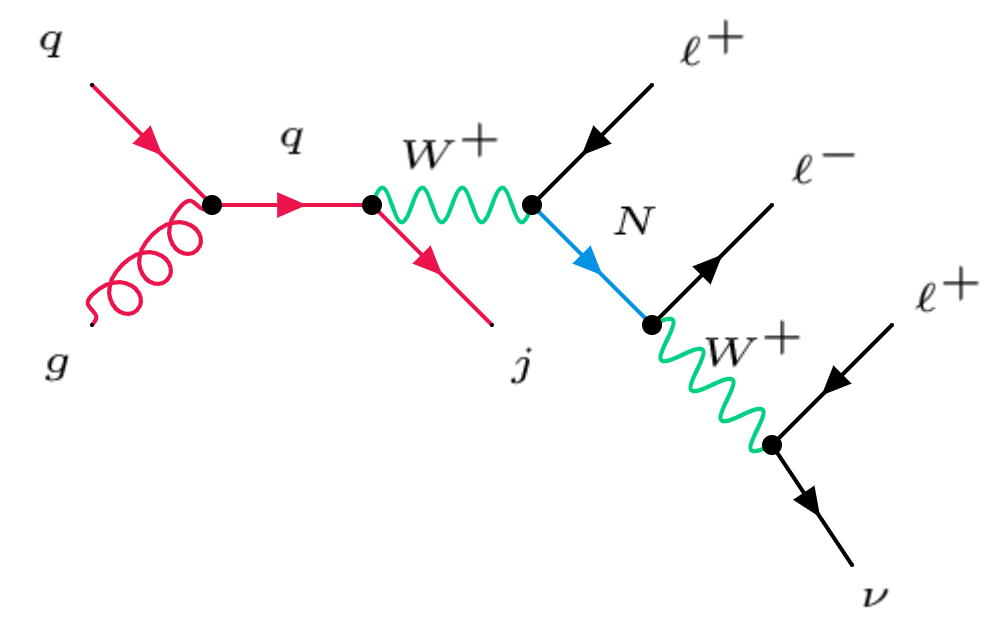}
  \includegraphics[width=0.29\textwidth]{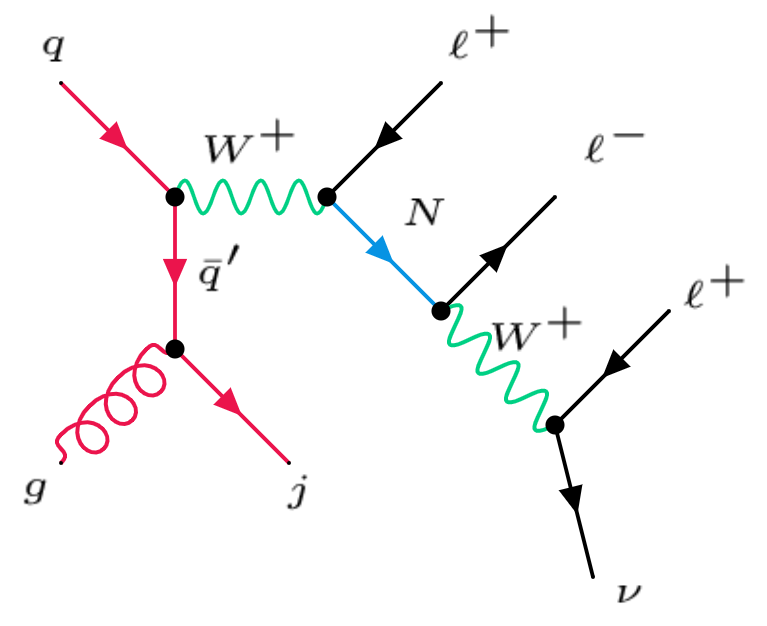}
  \caption{The trilepton plus MET signal of a heavy Dirac neutrino at the hadron colliders, without any jet (left panel) or with only one jet (middle and right panels).}
  \label{fig:diagram1}
\end{figure}

As seen in Eq.~(\ref{eqn:mnu}), the active neutrino masses are highly suppressed by $\mu_S$, and the Yukawa couplings $Y$ in Eq.~(\ref{eqn:Lagrangian}) are much larger than that in type-I seesaw with TeV-scale heavy neutrinos. This will enhance the on-shell production of heavy neutrinos at the LHC and future higher energy colliders. Following the pseudo-Dirac nature of heavy neutrinos, the same-sign dilepton signal $\ell_\alpha^\pm \ell_\beta^\pm W^\mp$ at the hadron colliders is highly suppressed by $\mu_S$. 
For $m_N \lesssim 1$ TeV, the primary signatures of heavy neutrinos at the hadron colliders are from the trilepton channel~\cite{Degrande:2016aje, Willenbrock:1985tj, Dicus:1991wj, Hessler:2014ssa, Ruiz:2017yyf}
\begin{eqnarray}
\label{eqn:trilepton}
pp & \to & \ell_\alpha^\pm  \overset{\textbf{\fontsize{2pt}{2pt}\selectfont(---)}}{N} \ \to \ \ell_\alpha^\pm \ell_\beta^\mp W^\pm \ \to \ \ell_\alpha^\pm \ell_\beta^\mp \ell_\gamma^\pm \nu \,. 
\end{eqnarray}
With the $W$ boson decaying leptonically, there are three charged leptons and significant MET from the neutrino in the final state. For simplicity, we have assumed that there is only one heavy neutrino involved in the collider signature analysis below, and all other heavy neutrinos are sufficiently heavy such that they are irrelevant. 
The corresponding leading-order parton level Feynman diagram is shown in the left panel of Figure~\ref{fig:diagram1}. Although the tiny neutrino masses are proportional to  $\mu_S$, the collider signatures in this paper do not depend on it. 
For the mass range of $m_N \gtrsim 1$ TeV, the vector-boson fusion (VBF) and gluon fusion processes become the dominant production channels~\cite{Datta:1993nm, Dev:2013wba, Alva:2014gxa, Degrande:2016aje}. For example, for the case of $m_{N}=500$ GeV, the cross section for DY production at the LHC 14 TeV is 63.3 fb when the 0-jet and 1-jet processes are combined, while the cross sections in the gluon fusion and VBF channels are only 17.9 fb and 15.0 fb, respectively. For simplicity, in this paper we study the mass range $m_N <1$ TeV and focus only on the DY production mode. 
 


The production in Eq.~(\ref{eqn:trilepton}) is mediated by the mixing $V_{\ell N}$ of $N$ with the active neutrino $\nu_\ell$, and we have assumed the heavy neutrino mass $m_N > 100$ GeV such that the decay $N \to \ell^- W^+$ is kinematically allowed. Then the cross section for the signal in Eq.~(\ref{eqn:trilepton}) can be written as 
\begin{eqnarray}
\sigma (pp \to \ell_\alpha^\pm \ell_\beta^\mp \ell_\gamma^\pm \nu) \simeq 
\sigma (pp \to \ell_\alpha^\pm \overset{\textbf{\fontsize{2pt}{2pt}\selectfont(---)}}{N}) \times {\rm BR} (\overset{\textbf{\fontsize{2pt}{2pt}\selectfont(---)}}{N} \to \ell_\beta^\mp W^\pm) \times {\rm BR} (W^\pm \to \ell_\gamma^\pm \nu) \,,
\end{eqnarray}
with ${\rm BR} (\overset{\textbf{\fontsize{2pt}{2pt}\selectfont(---)}}{N} \to \ell_\beta W^\pm)$ and ${\rm BR} (W^\pm \to \ell_\gamma^\pm \nu)$ the branching ratios (BRs) for the decays $\overset{\textbf{\fontsize{2pt}{2pt}\selectfont(---)}}{N} \to \ell_\beta^\mp W^\pm$ and $W^\pm \to \ell_\gamma^\pm \nu$, respectively. As the reconstruction of $\tau$ is highly non-trivial at hadron colliders, we will consider only the flavors $\alpha,\, \beta,\, \gamma = e,\;\mu$ for the collider analysis in this paper.  The tree-level differential cross section for the subprocess $ q q^\prime \to \ell N$ is~\cite{Datta:1993nm} 
\begin{eqnarray}
\frac{d \hat{ \sigma} (q q^\prime \to W^* \to \ell N)}{d \hat{t} } & = & \frac{\pi \alpha_W^2 |V_{\ell N}|^2} {12 \hat{s}^2} \frac{\hat{t} (\hat{t} - m_{N}^2) }{(\hat{s} - m_W^2)^2}\,,
\end{eqnarray}
where $\hat{s}$ and $\hat{t}$ are Mandelstam variables,  $m_W$ is the $W$ boson mass, $\alpha_W$ is the weak interaction constant defined as $\alpha_W \equiv g^2/4 \pi$ with $g$ the gauge coupling for the SM gauge interaction $SU(2)_L$.  
The heavy neutrino $N$ can decay into the SM particles via the two-body channels $N \to \ell^- W^+$, $\nu_\ell Z$, $\nu_\ell h$, and the corresponding partial decay widths are given by~\cite{Buchmuller:1991tu, Pilaftsis:1991ug} 
\begin{eqnarray}
\label{eqn:width1}
\Gamma(N \to \ell^- W^+) & \ = \ & \frac{\alpha_W |V_{\ell N}|^2}{16}  \frac{m_N^3}{m_W^2} \left(1-\frac{m_W^2}{m_N^2}\right)^2 \left(1+\frac{2m_W^2}{m_N^2}\right) \,, \\
\Gamma(N \to \nu_\ell Z) & \ = \ & \frac{\alpha_W |V_{\ell N}|^2}{32 \cos^2\theta_W}  \frac{m_N^3}{m_Z^2} \left(1-\frac{m_Z^2}{m_N^2}\right)^2
\left(1+\frac{2m_Z^2}{m_N^2}\right) \,, \\
\label{eqn:width3}
\Gamma(N \to \nu_\ell h) & \ = \ & \frac{\alpha_W |V_{\ell N}|^2}{32}  \frac{m_N^3}{m_W^2}\left(1-\frac{m_h^2}{m_N^2}\right)^2 \,,
\label{widths}
\end{eqnarray}
where $\theta_W$ is the Weinberg mixing angle, and $m_{Z}$ and $m_h$ are the masses of the SM $Z$ and Higgs bosons, respectively. With these decay widths, it is straightforward to calculate ${\rm BR} (N \to l^- W^+)$. 
Furthermore, for the parameter space of $m_N > 100$ GeV and $|V_{\ell N}|^2 \gtrsim 10^{-5}$ of interest in this paper (cf. Fig.~\ref{fig:Bln}), the corresponding lifetime of heavy neutrino $\tau_N \lesssim 10^{-9}$ cm, making the heavy neutrino short-lived enough at the high-energy colliders and the decay products prompt signals.

If we have one or two jets in the trilepton events, the production cross sections at the LHC and future higher energy colliders are also significant~\cite{Das:2014jxa, Das:2015toa, Degrande:2016aje, Das:2016hof, Pascoli:2018rsg, Ruiz:2015zca}. At the parton level, there are only leptons in the final state at the leading-order, therefore the extra jets are from initial state radiation, as shown in the middle and right panels of Fig.~\ref{fig:diagram1}. In this case, the $qg$ process is also important, and it turns out that the cross section for $pp \to N \ell^+ (\bar{N} \ell^-) + 1j$ is only two to four times smaller than that for the leading-order process without any jet at the LHC 14 TeV. At the HE-LHC 27 TeV and future 100 TeV colliders the trilepton events with one or two jets are more important than at the LHC. More details can be found in Section~\ref{sec:analysis} below.

\section{Collider Analysis}
\label{sec:analysis}

In this section we perform a detailed collider analysis on both the pure leptonic final states of the signature in Eq.~(\ref{eqn:trilepton}) and the case with extra jets 
at $\sqrt{s} = $ 14 TeV, 27 TeV and 100 TeV. In this study, we take the heavy neutrino mass range from 100 GeV to 1 TeV, for which the process in Eq.~(\ref{eqn:trilepton}) is the dominant channel. 

\subsection{MC event generation}
\label{sec:event}

All the MC event samples are generated using {\tt MadGraph5\_aMC@NLO v2.8.2}~\cite{Alwall:2014hca}, and then passed to {\tt PYTHIA8}~\cite{Sjostrand:2014zea} for parton showering, fragmentation, and hadronization. The detector response is evaluated by using {\tt DELPHES 3}~\cite{deFavereau:2013fsa} with an ATLAS detector card. In both background and signal processes the parton distribution function (PDF) data sets {\tt NN23LO1 PDF}~\cite{Ball:2012cx,Ball:2013hta} are chosen, and the MLM~\cite{Mangano:2001xp,Mangano:2006rw} merging/matching procedure is adopted with {\tt xqcut} set to be 40 GeV. {\tt FastJet}~\cite{Cacciari:2011ma} is integrated by {\tt DELPHES} for jet reconstruction, and the anti-$k_T$ algorithm~\cite{Cacciari:2008gp} is adopted with the parameter {\tt R} set to be 0.6 and {\tt JetPTMin} set to be 20 GeV.
The decay tables of heavy neutrino $N$ and the $Z$ and $W$ bosons are computed by {\tt MadGraph5}. The decay of $\tau$ lepton generally produces lots of jets in the detector, which makes the $\tau$ lepton signals suffer a severe contamination from the huge QCD backgrounds. Therefore, as mentioned above, in this paper we will consider only electrons and muons in the leptonic final states as our signal events. 

\begin{table}[!t]
\centering
\caption{  Geometrical acceptance cuts of electrons, muons and jets adopted in our data simulations at $\sqrt{s} = 14$ TeV, 27 TeV and 100 TeV. In addition, for the leading lepton we take $p^{\ell,\,{\rm leading}}_{T} > $ 20 GeV.}
\label{tab:acc}
\vspace{5pt}
 \begin{tabular}{|c | c c | c c | c c |} 
 \hline
  \multirow{2}*{$\sqrt{s}$  (TeV)}   &  \multicolumn{2}{c|}{electron} & \multicolumn{2}{c|}{muon} &  \multicolumn{2}{c|}{jet}\\
  \cline{2-7}
            & $p_{T}$ (GeV) & $|\eta|$ &  $p_{T}$ (GeV) & $|\eta|$ & $p_{T}$ (GeV) & $|\eta|$\\
  \hline
14  & $>$10 & $<$ 2.47& $>$10 & $<$ 2.7 & $>$ 20 & $<$ 4.5\\
27  & $>$10 & $<$ 2.47& $>$10 & $<$ 2.7 & $>$ 30 & $<$ 4.5\\
100	& $>$15 & $<$ 2.47& $>$15 & $<$ 2.7 & $>$ 45 & $<$ 4.5\\
 \hline
\end{tabular}
\end{table}

At the detector level,  the visible physical objects include jets, electrons and muons for our case, which can be reconstructed by their momenta $p_T$, pseudorapidities $\eta$ and azimuthal angles $\phi$. The MET $E_T^{\rm miss}$ can be reconstructed from all the visible objects via $E_T^{\rm miss} = |-\sum_{v_i} {\stackrel{\rightarrow}{p_{T}}}( v_i)|$, with $\stackrel{\rightarrow}{p_T}(v_i)$ the transverse momentum of the $i$th visible object.  We will use the reconstructed jets, charged leptons, and MET to perform our signal and background analysis below. 
It has been examined that the total inclusive cross sections below are independent of the matching parameter. More explicitly, at the parton level we adopt the basic acceptance cuts for both signal and background events in Table~\ref{tab:acc}. At $\sqrt{s} = $14 TeV and 27 TeV, the  cuts are the same as those used in the ATLAS lepton analyses at $\sqrt{s}$ = 13 TeV in Ref.~\cite{Aaboud:2018jiw}. At $\sqrt{s} = $100 TeV, the $p_{T}$ cuts on charged leptons and jets are relatively larger than that at 14 TeV and 27 TeV to remove the soft processes. For concreteness, we set $p_T >$ 15 GeV for electrons and muons, and $p_T>$ 45 GeV for jets. In addition, we require at least one lepton has the transverse momentum larger than 20 GeV, i.e. $p^{\ell,\,{\rm leading}}_{T} > $ 20 GeV. The cuts on $|\eta|$ are the same, assuming that the detector geometry is identical to ATLAS. 

\begin{figure}[t]
  \centering
  \includegraphics[width=0.32\textwidth]{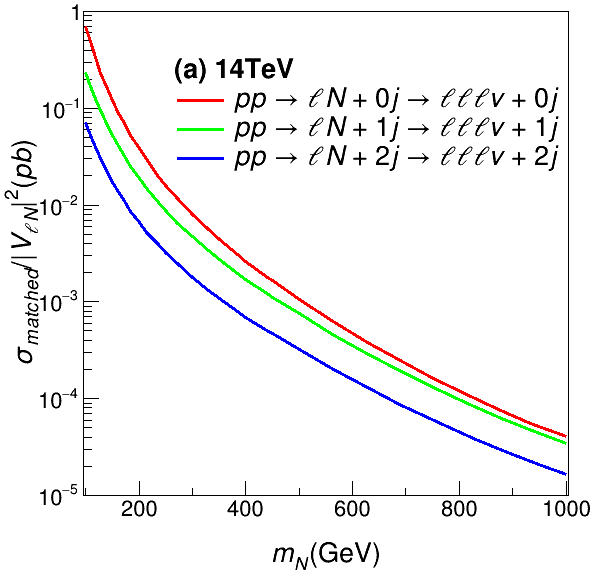}
  \includegraphics[width=0.32\textwidth]{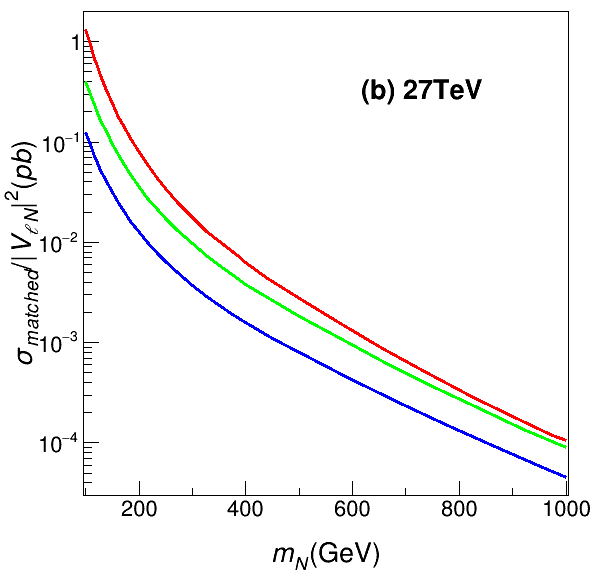}
  \includegraphics[width=0.32\textwidth]{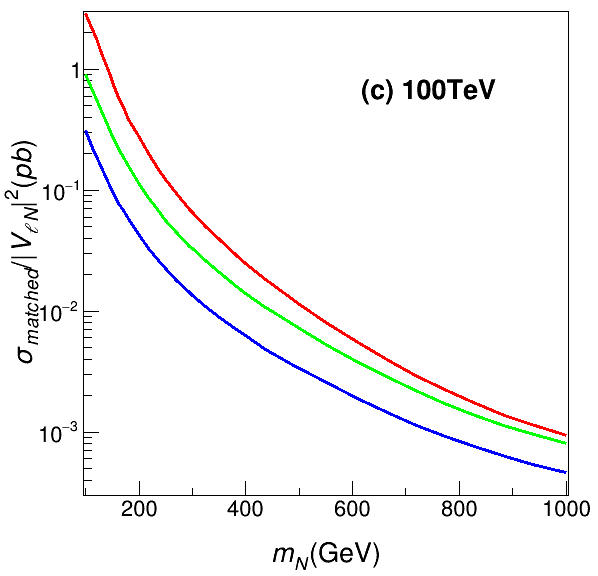}
  \caption{Cross sections of trilepton signal process with $m_N=100$ GeV to 1 TeV at $\sqrt{s}=$ 14 TeV (left), 27 TeV (middle) and 100 TeV (right), with $\ell=\mu$.  Exclusive cross sections for the processes with 0 jet (red), 1 jet (green) and 2 jets (blue) are shown for comparison. The acceptance cuts in Table~\ref{tab:acc} have been applied. }
  \label{fig:xs}
\end{figure}

The signal is generated based on the model file {\tt HeavyN} for Dirac neutrinos~\cite{Christensen:2008py,Alloul:2013bka,Ruiz:2020cjx, Alva:2014gxa,Degrande:2016aje, Pascoli:2018heg}. As stated in Section~\ref{sec:model}, for simplicity we have assumed that there is only one heavy neutrino $N$ with mass at the electroweak scale, while the other two heavy neutrinos are much heavier and decouple from our analysis at the high-energy colliders. The exclusive cross sections for the trilepton signal processes in Eq.~(\ref{eqn:trilepton}) at $\sqrt{s}=$ 14 TeV, 27 TeV and 100 TeV are shown in the left, middle and right panels of Fig.~\ref{fig:xs}, respectively, as function of the heavy neutrino mass $m_N$. For the sake of concreteness, we have set the lepton flavor $\ell = \mu$. For the exclusive processes, we have included also the $pp \to \ell N + j$ and $pp \to \ell N + 2 j$ processes. As shown in the middle and right panels of Fig.~\ref{fig:diagram1}, there are not only $q \bar{q}^\prime$ processes, but also $gq$ (or $g \bar{q}$) and $gg$ processes. The exclusive cross sections for signals with 0, 1 and 2 jets are shown in Fig.~\ref{fig:xs}, respectively, by the red, green and blue curves, and the acceptance cuts in Table~\ref{tab:acc} have been applied. It is noteworthy that for a fixed $m_N$ the cross sections for the processes with one jet are roughly three times smaller than those without any jet. A similar suppression factor can be observed when the cross sections with two jets are compared with those with only one jet. Given the exclusive cross sections in Fig.~\ref{fig:xs}, the signal production cross sections can be modelled well by the first three exclusive data samples, and it is safe to neglect events with higher jet multiplicity $n_j\geq 3$ in the following study.


\begin{table}[!t]
\centering
\caption{All possible trilepton final states in charge and flavor spaces, assuming either $|V_{eN}|\neq0$ or $|V_{\mu N}|\neq0$. }
\label{tab:lll}
\vspace{5pt}
\begin{tabular}{|c|c|c|}
\hline
mixing & trilepton states  & signs ($\pm\mp\pm$) \\ \hline
\multirow{2}*{$V_{eN}$} & $eee$ & $e^\pm e^\mp e^\pm$ \\ \cline{2-3}
& $ee\mu$ &  $e^\pm e^\mp \mu^\pm $ \\  \hline
\multirow{2}*{$V_{\mu N}$} & $\mu\mu e$ & $\mu^\pm \mu^\mp e^\pm $ \\ \cline{2-3}
& $\mu\mu\mu$ &  $ \mu^\pm \mu^\mp \mu^\pm$ \\ \hline
\end{tabular}
\end{table}

For simplicity, in the following simulations, we assume there is only either $V_{eN}\neq0$ or $V_{\mu N}\neq0$ in the neutrino sector, which induces the combinations of charged lepton flavors $eee$ and $ee\mu$ or $\mu\mu e$ and $\mu\mu\mu$, with the first, second and third one corresponding, respectively, to the flavor indices $\alpha$, $\beta$ and $\gamma$ in Eq.~(\ref{eqn:trilepton}).  In the charge space, the three leptons of an event can be either $+-+$ or $-+-$. All the resultant trilepton states are tabulated in Table~\ref{tab:lll}. Our trilepton observables below will be constructed in both the charge and flavor spaces. Note that the flavor combinations $\ell_\alpha^\pm \ell_\beta^\mp \ell_\alpha^\pm$ ($\alpha \neq \beta$) imply that we have both nonzero $V_{eN}$ and $V_{\mu N}$, which is a clear signature of lepton flavor violation (LFV). For instance, the states $e^\pm \mu^\mp e^\pm$ can originate from the production and decay chain:
\begin{eqnarray}
pp & \to & e^\pm \overset{\textbf{\fontsize{2pt}{2pt}\selectfont(---)}}{N} 
\ \to \ e^\pm \mu^\mp W^\pm \ \to \ e^\pm \mu^\mp e^\pm \nu \,,
\end{eqnarray}
where the process $pp\to eN$ is induced by the neutrino mixing $V_{eN}$ while the decay $N \to \mu W$ is from the mixing $V_{\mu N}$. Such a LFV process is clearly absent in the SM. 

The main background events for our signal are from $ p p \to Z W^\pm$ process in the SM, where both $Z$ and $W^\pm$ bosons decay into leptonic final states, i.e.
\begin{eqnarray}
pp & \to & Z W^\pm \ \to \ \ell_1^\pm \ell_2^\mp \ell_3^\pm \nu \,,
\label{bgtl}
\end{eqnarray}
with potentially one or two additional jets. The SM backgrounds have the same particles in the final state as the signal events. 
To select the signal events, we demand the following pre-selection conditions for every event: 
\begin{itemize}
\item The total number of energetic charged leptons is exactly 3, i.e. $n_e + n_\mu = 3$.
\item In order to suppress the background events from the low mass DY process and hadronic decays, the invariant mass of the two leading leptons should be larger than 12 GeV, i.e. $m_{\ell_1 \ell_2}$ $>$ 12 GeV.
\item The number of jets is not larger than 2, i.e. $n_j \leq 2$.
\item The missing transverse momentum $E_T^{\rm miss} = |-\sum_{v_i}\stackrel{\rightarrow}{p_T}(v_i)|$ is larger than 20 GeV, i.e. $E^{\rm miss}_{T} > 20$ GeV. 
\end{itemize}

The SM backgrounds receive also contributions from the off-shell processes $pp \to Z^\ast W,\, \gamma^\ast W \to \ell \ell \ell \nu$~\cite{Pascoli:2018heg}, with the photon process interfering with the $Z$-boson mediated diagrams. To estimate the cross section for the off-shell processes and interference terms, we first obtain the cross section for the purely on-shell $ZW \to \ell\ell\ell\nu$ process, and then the total cross section for the same final state, combining all the contributions from the on-shell $ZW$, off-shell $Z^\ast W$, virtual photon $\gamma^\ast W$ and the interference terms. Subtracting the on-shell $ZW$ contribution from the total cross section, we obtain the off-shell and interference contributions to the backgrounds. As a demonstration, the cross sections for the on-shell and off-shell plus interference contributions with $\ell=\mu$ are shown in the second and third rows of Table~\ref{tab:xs:bkg}, where we have applied the geometrical acceptance cuts in Table~\ref{tab:acc} and the pre-selection cuts above. As seen in this table, for the 0-jet and 1-jet processes, the off-shell plus interference contributions are only roughly 6\% and 4\% of the cross sections for the corresponding on-shell processes.

In addition, the four-lepton production $4\ell$ contributes to the backgrounds, if one of the charged leptons is missed by  the detectors~\cite{Pascoli:2018heg}. However, after the geometrical acceptance cuts and pre-selection cuts in particular the missing energy cut, such a process can be effectively removed. As demonstrated in the last row of Table~\ref{tab:xs:bkg}, for the 0-jet and 1-jet cases, the four-lepton process contributes only roughly 5\% and 6\% to the backgrounds, respectively. Taking into account the contributions of the sub-dominant off-shell and four-lepton processes to the SM backgrounds, we scale the cross sections of the $ZW^\pm$ background, e.g. by, respectively, a factor of 1.11 and 1.10 for the cases of 0-jet and 1-jet with $\ell = \mu$ in Table~\ref{tab:xs:bkg}. 


\begin{table}[!t]
\centering
\caption{Cross sections of the on-shell $ZW$, off-shell plus interference and four-lepton $4\ell$ background processes after geometrical acceptance cuts presented in Table~\ref{tab:acc} and the pre-selections cuts, with $\ell = \mu$.}
    \label{tab:xs:bkg}
\vspace{5pt}
 \begin{tabular}{|c | c | c |} 
 \hline
 \multirow{2}*{process} & \multicolumn{2}{c|}{cross section (fb)} \\ \cline{2-3}
    &  0-jet & 1-jet \\
  \hline
$ZW \rightarrow \ell\ell\ell\nu$  & 29.10 & 20.50 \\ \hline
$\ell\ell\ell\nu$ & \multirow{2}*{1.65} & \multirow{2}*{0.84} \\ 
(off-shell + interference) && \\ \hline
$4\ell$	                  & 1.56 & 1.25\\
 \hline
\end{tabular}
\end{table}

\begin{figure}[t]
  \centering
  \includegraphics[width=0.5\textwidth]{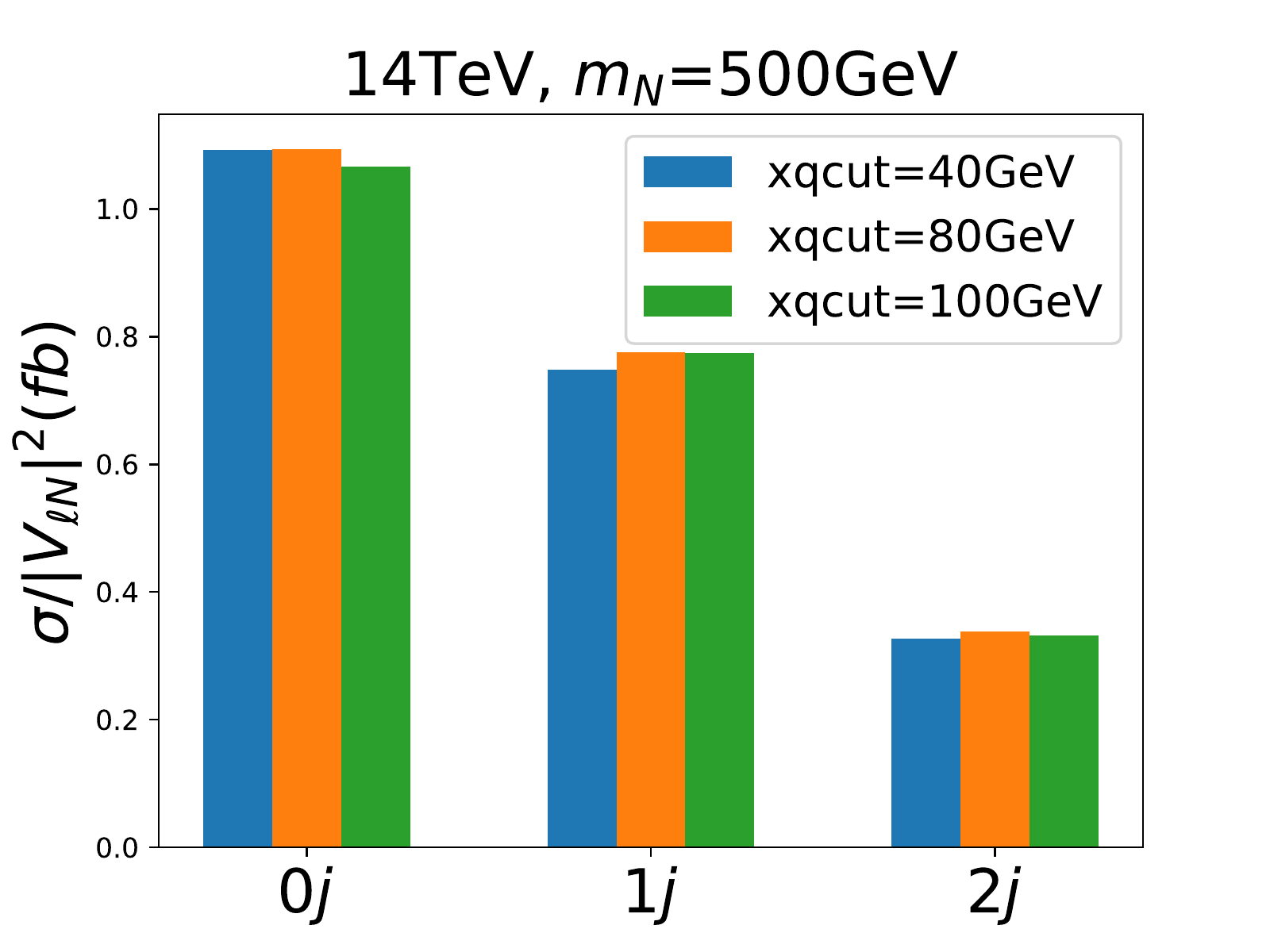}
  \caption{Signal cross sections for final states with 0, 1, and 2 jets at $\sqrt{s}=$ 14 TeV, with $m_N = 500$ GeV. The parameter $xqcut$ is taken to be 40 GeV, 80 GeV and 100 GeV for comparison.}
  \label{fig:xqcut}
\end{figure}

We use {\tt MadGraph5} to generate the matrix elements of the signal processes $p p \to \ell N$, $p p \to \ell N + j$, and $p p \to  \ell N + 2j $ and the background processes $p p \to W Z$, $p p \to W Z + j$, and $p p \to W Z+ 2j $. The MLM merging/matching procedure is adopted to produce $10^4$ events as an inclusive data sample, and three values are taken for the jet parameter $xqcut$, i.e. $xqcut = 40$ GeV, 80 GeV, and 100 GeV. As an explicit example, the cross sections $\sigma/|V_{\ell N}|^2$ for the signal of $m_N = 500$ GeV with 0, 1 and 2 jets at $\sqrt{s}=14$ TeV for the three $xqcut$ values are shown in Fig.~\ref{fig:xqcut}. As clearly seen in this figure, the signal cross sections are almost the same, differing by $\sim 1 \%$ for the $xqcut$ values above. For the matched cross sections for the signal processes with 0, 1 and 2 jets at $\sqrt{s}=14$ TeV, 27 TeV and 100 TeV in Fig.~\ref{fig:xs}, we have taken $xqcut=40$ GeV. 


To demonstrate the task of background and signal event discrimination, we present in Table~\ref{tab:xs} the cross sections of signal events with $m_N= 150$ GeV and $m_N=500$ GeV, and that of the SM background process $pp \to Z W^\pm + \textrm{jets}$ at $\sqrt{s}=$ 14 TeV, 27 TeV and 100 TeV. All the signal cross sections are normalized to $|V_{\ell N}|^2=1$. The exclusive cross sections for $0$, $1$, and $2$ jets are all presented in Table \ref{tab:xs}, and we have applied the acceptance cuts in Table~\ref{tab:acc}. It can be observed in Table~\ref{tab:xs} that when $m_N$ is small, say 150 GeV, the cross sections of signal can be comparable with that of the SM background processes at $\sqrt{s} = 14$ TeV, 27 TeV and 100 TeV. When $m_N$ is large, say 500 GeV, the signal cross sections can be roughly 50 times smaller than that for the SM backgrounds at 14 TeV, and down to roughly 30 times smaller at the 100 TeV collider.







\begin{table}[!t]
\centering
\caption{  Exclusive cross sections of trilepton signal process in Eq.~(\ref{eqn:trilepton}) with $m_N = 150$ GeV or 500 GeV and the dominant SM background in Eq.~(\ref{bgtl}) at $\sqrt{s} =14 $ TeV, 27 TeV and 100 TeV, with 0, 1 or 2 jets. We have taken $\ell=\mu$ and adopted the acceptance cuts in Tab.~\ref{tab:acc}. Note that the cross sections for the SM backgrounds with 0 and 1 jet at 14 TeV are different from those in Table~\ref{tab:xs:bkg}, as the cuts are different.
}
\label{tab:xs}
\vspace{5pt}
\scalebox{0.925}{
 \begin{tabular}{|c | c c c | c c c | c c c|} 
 \hline
  \multirow{3}*{$\sqrt{s}$  (TeV)}   &  \multicolumn{3}{c|}{$m_{N}$ = 150 GeV} & \multicolumn{3}{c|}{$m_{N}$ = 500 GeV} &  \multicolumn{3}{c|}{SM background}\\
  \cline{2-10}
           &  \multicolumn{6}{c|}{$\sigma_{\rm matched}$/$|V_{\ell N}|^2$ (pb)} & \multicolumn{3}{c|}{$\sigma_{\rm matched}$ (pb)}\\
  \cline{2-10}
            & 0-jet & 1-jet & 2-jet & 0-jet & 1-jet & 2-jet & 0-jet & 1-jet & 2-jet \\
  \hline
14 & 0.121 & 0.0521 & 0.0172 & 0.00107 & 0.00075 & 0.00032 & 0.0561 & 0.0392 & 0.0214 \\
27 & 0.241 & 0.0947 & 0.0313 & 0.00275 & 0.00184 & 0.000795 & 0.107 & 0.0842 & 0.0475 \\
100 & 0.804 & 0.269 & 0.0923 & 0.0119 & 0.00705 & 0.00321 & 0.346 & 0.327 & 0.183 \\ \hline

\end{tabular}}
\end{table}

\subsection{Feature Observables}
\label{sect:variables}

In order to enhance the efficiency of discriminating the background and signal events, in this subsection we define some feature observables. To reconstruct these observables from charged leptons in the final state, it is crucial to identify their origins in terms of the background or signal hypothesis. To this end, we use the background $W^- Z$ and the signal $e^- N$ as explicit examples to describe how to identify the origins of charged leptons in our reconstruction algorithm.
\begin{itemize}
\item For the $W^- Z \to e^{-}e^{+} e^{-} + E_T^{\rm miss}$ background events, it is known that the charged lepton with positive charge, i.e. $e^{+}$, must come from the $Z$ boson, labeled as $\ell_2^+$.  To determine the second lepton $e^-$ from $Z$ decay, there is a two-fold ambiguity. In order to choose it out, we demand that the lepton $\ell_1^-$ has the minimal $|m_{\ell_2^+ \ell_1^-} - M_Z|$. Then the last lepton $e^-$ is from $W$ boson decay and can be labelled as $\ell_3^-$. This would fix the background kinematics $W^- Z \to \ell_1^- \ell_2^+ \ell_3^- \bar\nu = e^- e^+ e^- \bar\nu$. This can be easily generalized to the cases of $\mu^- \mu^+ \mu^- + E_T^{\rm miss}$ and the charge conjugate $W^+ Z$ background.



\item For the background mode $e^{-}e^{+}\mu^{-} + E_T^{\rm miss}$, the determination of kinematics is rather trivial: the opposite-sign same-flavor lepton pair is from the $Z$ boson decay, while the third one is from the $W$ boson, i.e. $W^- Z \to \mu^- \bar\nu e^+ e^-$. It is equally trivial for the background $\mu^- \mu^+ e^- \bar\nu$. 
\end{itemize}
Generally speaking, in the background hypothesis, we need to first reconstruct the $Z$ boson from a pair of charged leptons $\ell_1^\pm \ell_2^\mp$, and then the kinematic properties of $W$ boson can be reconstructed from the third charged lepton $\ell_3^\pm$ and the missing energy. On the other hand, for the signal hypothesis, the kinematic variables of the heavy neutrino are related to $\ell_2^\mp$, $\ell_3^\pm$ and the missing energy. 
Note that {\tt DELPHES} is unable to handle fakes for charge identification. In the detector, bremsstrahlung and straight-like tracks produced at high $p_T$ are the main sources of lepton charge mis-identification. The mis-identification portion is only about 0.3\%, which will not significantly affect our results~\cite{FernandezPretel:2791516}.

After identifying the origins of these leptons in terms of background and signal hypotheses, we can reconstruct the feature observables, which are divided into three categories, i.e. global kinematic observables, observables for background event vetoing, and observables favoring signal events. Since there are at least 4 particles in the phase space of our signal and background events, we need at least 12-dimension phase space, or equally at least 12 independent observables, to describe the collider events. Since the longitudinal component of the momentum of neutrino in the final state is missing, we can use the hypothesis that missing energy is from $W$ boson decay and derive its expected value.

\begin{figure}[t]
  \centering
  \includegraphics[width=0.32\textwidth]{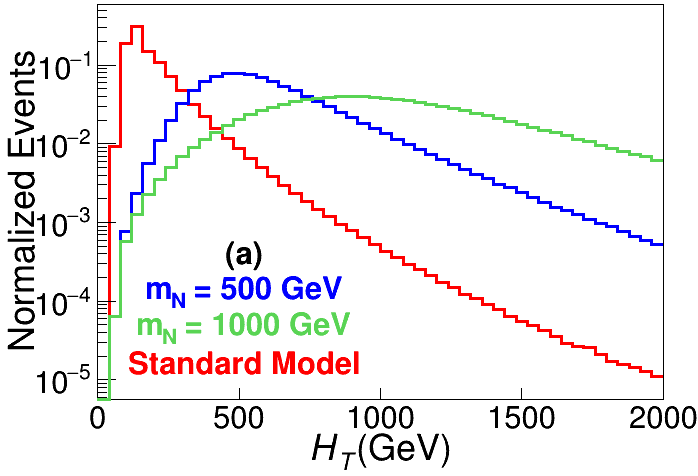}
  \includegraphics[width=0.32\textwidth]{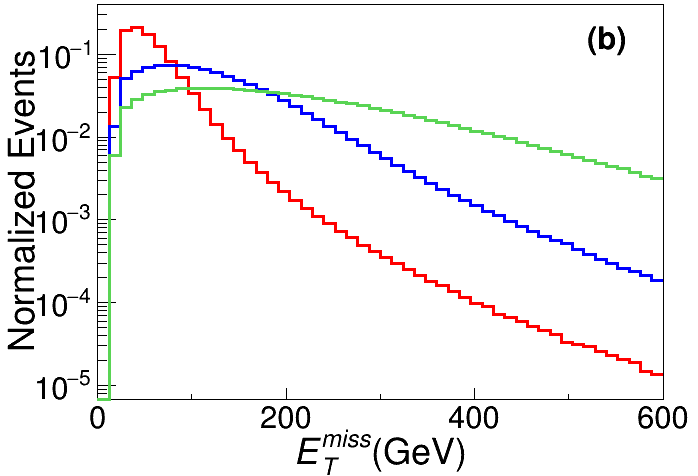}
  \includegraphics[width=0.32\textwidth]{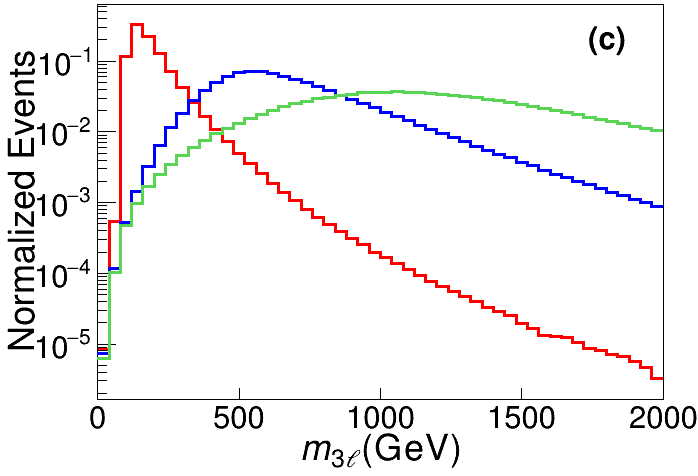}
  \caption{Distributions of global observables $H_T$ (left), $E_T^{\rm miss}$ (middle) and $m_{3\ell}$ (right) of SM backgrounds (red) and heavy neutrino signals with $m_N = 500$ GeV (blue) and 1000 GeV (green) at $\sqrt{s}=$ 14 TeV. The distributions are normalised so that their integrals are equal to 1.}
  \label{fig:var_global}
\end{figure}

Let us first consider the events without any jet, and list the definitions of the global kinematic observables for  background and signal discrimination. For illustration purpose, the distributions of SM backgrounds and signals with $m_N = 500$ GeV and 1000 GeV at $\sqrt{s} =14$ TeV are presented in Fig.~\ref{fig:var_global} as the red, blue and green lines, respectively. 
\begin{itemize}
\item The scalar sum of transverse momenta $H_T=\sum_{v_i}|\stackrel{\rightarrow}{p_T}(v_i)|$, where  $v_i$ runs over all the reconstructed visible particles. This quantity is related to the mass parameter $m_N$ for signal events; for background events, it related to the transverse momenta of the $W$ and $Z$ bosons. The distributions for the SM backgrounds and signals are shown in the left panel of Fig.~\ref{fig:var_global}. 
\item The missing transverse momentum $E_T^{\rm miss} = |-\sum_{v_i}\stackrel{\rightarrow}{p_T}(v_i)|$. The corresponding distributions for backgrounds and signals are presented in the middle panel of Fig.~\ref{fig:var_global}. \item The invariant mass $m_{3 \ell}$ of trilepton final state.   This quantity is related to the thresholds of the signal and background processes. As seen in the right panel of Fig.~\ref{fig:var_global}, the distributions of $m_{3\ell}$ are significantly different for the backgrounds and signals. 
\end{itemize}

\begin{figure}[t]
  \centering
  \includegraphics[width=0.32\textwidth]{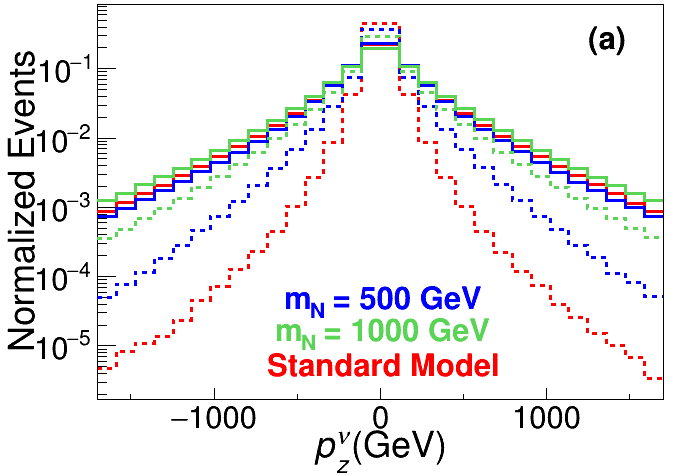}
  \includegraphics[width=0.32\textwidth]{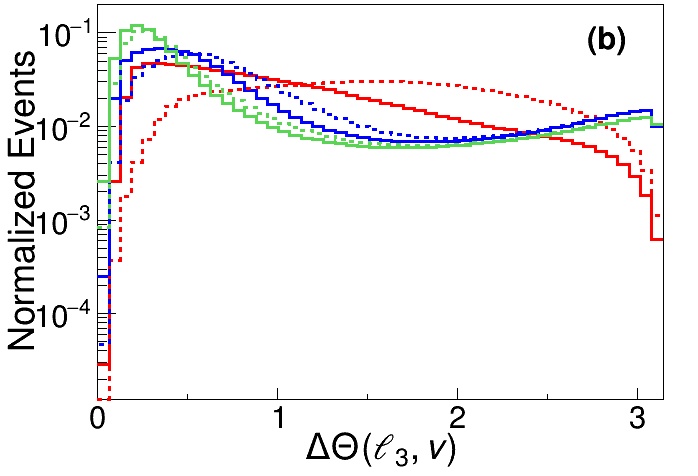}
  \includegraphics[width=0.32\textwidth]{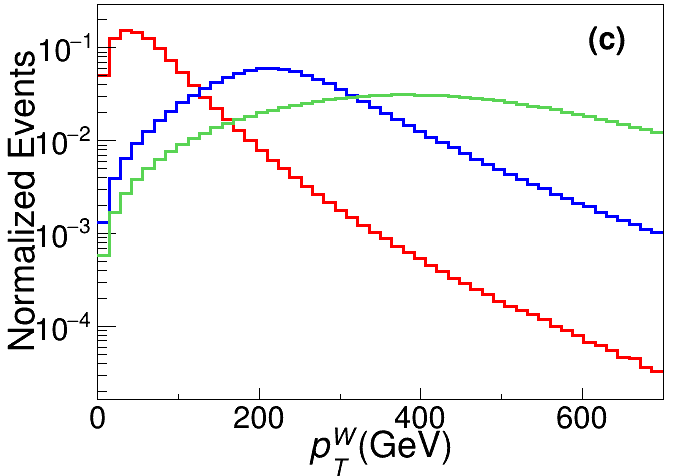}
  \includegraphics[width=0.32\textwidth]{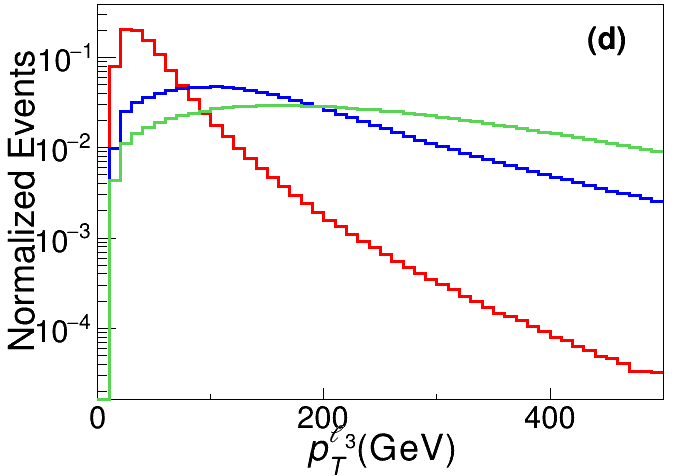}
  \includegraphics[width=0.32\textwidth]{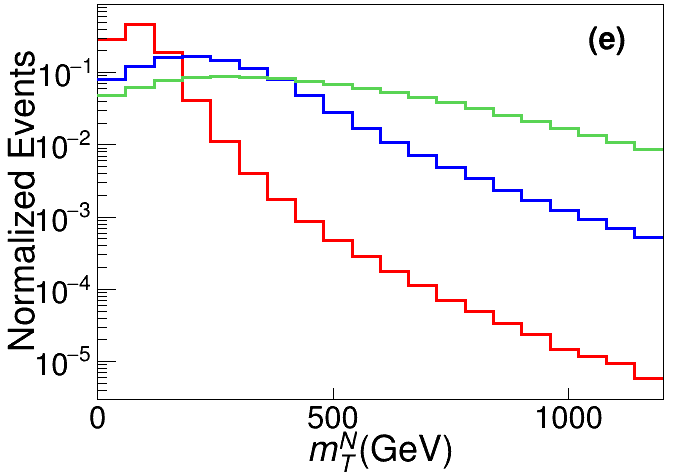}
  \includegraphics[width=0.32\textwidth]{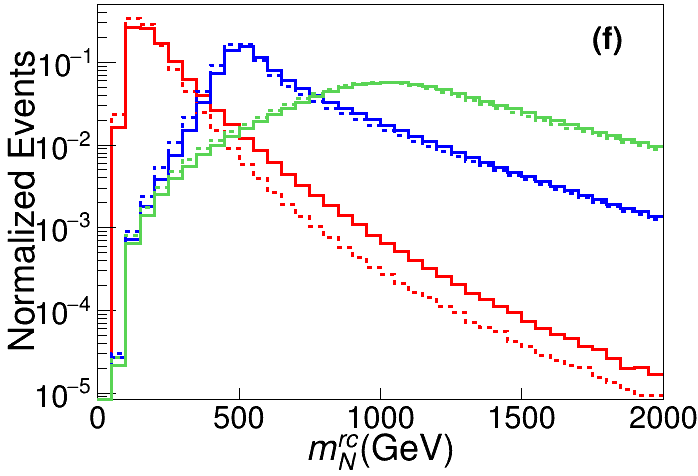}

  \caption{Distributions of signal related observables $p_z^\nu$ (upper left), $\Delta \Theta (\ell_3,\, \nu)$ (upper middle), $p_T^W$ (upper right), $p_T^{\ell_3}$ (lower left), $m_T^N$ (lower middle) and $m_N^{\rm rc}$ (lower right) of SM backgrounds (red) and heavy neutrino signals with $m_N = 500$ GeV (blue) and 1000 GeV (green) at $\sqrt{s}=$ 14 TeV. As the variable $p_z^\nu$ is two-folded, the solid lines in the upper left, upper middle and lower right panels correspond to the solution 0 (larger absolute value of $p_z^{\nu}$) and the dash lines to the solution 1 (smaller absolute value of $p_z^{\nu}$). The distributions are normalised so that their integrals are equal to 1.  }
  \label{fig:var_sig}
\end{figure}

Six feature observables are introduced to describe the phase space in favor of the signal events, with the distributions for backgrounds and signals shown in Fig.~\ref{fig:var_sig}.
\begin{itemize}
\item The two-fold longitudinal momentum solutions $p^{\nu,i}_z$ (with $i=0,1$) of the light neutrino in the final state. In our analysis, the transverse momentum of light neutrino is defined by $p_T(\nu)=E^{\rm miss}_T=|-\sum_{v_i}\stackrel{\rightarrow}{p_T}(v_i)|$, where $v_i$ denotes charged leptons and jets.  Since the light neutrino masses $m_{\nu}\simeq 0$, we have $E^2_\nu=p^2_T(\nu)+p^2_z(\nu)$, where $E_\nu$ and $p_z(\nu)$ are the energy and longitudinal momentum of light neutrino, respectively~\cite{ATLAS:2012byx}.  
For the hypothesis that light neutrino is from the on-shell $W$ boson decay, we can have the relation $(p_{\ell}+p_\nu)^2=m^2_W$, where $p_{\ell}$ is the four-momentum of the charged lepton from $W$ decay. Then we can have an quadratic equation for $p_z(\nu)$, which yields two solutions $p_z^{\nu,0}$ and $p_z^{\nu,1}$, with $p_z^{\nu,0}$ labeling the one with the larger absolute value. More details can be found in the Appendix. 
The distributions of $p_z^{\nu,0}$ and  $p_z^{\nu,1}$ are shown, respectively, as the solid and dashed lines in the upper left panel of Fig.~\ref{fig:var_sig}.

\item The angular separation $\Delta\Theta (\ell_3,\nu)$ between the charged lepton $\ell_3$ and the neutrino from $W$ boson decay. As the longitudinal momentum of neutrino has a two-fold ambiguity, the separation $\Delta\Theta(\ell_3,\nu)$ is also two-folded, as seen in the upper middle panel of Fig.~\ref{fig:var_sig}. When the heavy neutrino mass $m_N$ is large, the daughter $W$ boson will be highly boosted, and the charged lepton and neutrino from its decay will be highly collimated. Therefore the separation $\Delta\Theta (\ell_3,\nu)$ can be viewed as a measurement of the Lorentz boost factor of the $W$ boson from $N$ decay.

\item The transverse momentum $p_{T}^{W}$ of the reconstructed $W$ boson. The corresponding distributions are presented in the upper right panel of Fig.~\ref{fig:var_sig}. Obviously, when $m_N$ is large, $p_{T}^{W}$ tends also to be large.

\item The transverse momentum $p_{T}^{\ell_{3}}$ of the lepton $\ell_{3}$ originating from $W$ boson decay. As seen in the lower left panel of Fig.~\ref{fig:var_sig}, $p_T^{\ell_3}$ tends to be large for the signal. 
\item The transverse mass $m_T^N$ of the heavy neutrino.  The proper mass $m_N$ of $N$ can not be fully reconstructed due to the missing information of longitudinal momentum of the neutrino in the final state. For each event, once the lepton $\ell_1$ is identified, the transverse mass can be  defined via 
\begin{eqnarray}
m_{T}^N=\sqrt{2 p_T^{\ell_{2}\ell_{3}} E_T^{\rm miss}(1-\cos{\Delta \phi ({\ell_{2}\ell_{3}, E_T^{\rm miss}}}))} \,,
\end{eqnarray}
where $p_T^{\ell_2 \ell_3}$ is the transverse momentum of the $\ell_2 \ell_3$ subsystem, and $\Delta \phi  (\ell_{2}\ell_{3}, E_T^{\rm miss})$ denotes the azimuthal angle between the $\ell_{2} \ell_{3}$ subsystem and the MET $E_T^{\rm miss}$. The distributions of $m_T^N$ are shown in the lower middle panel of Fig.~\ref{fig:var_sig}. 

\item Two reconstructed heavy neutrino masses $m_N^{{\rm rc},i}$ (with $i=0,\,1$), which has a two-fold ambiguity as a result of the two solutions $p_z^{\nu,0}$ and $p_z^{\nu,1}$. For signal events, one of them must be close to the real $m_N$. While for background events, neither of them is expected to be condensed in the mass window of $m_N$. As seen in the lower right panel of Fig.~\ref{fig:var_sig}, $m_N^{\rm rc}$ can well distinguish the signal from backgrounds. 
\end{itemize}

\begin{figure}[t]
  \centering
  \includegraphics[width=0.32\textwidth]{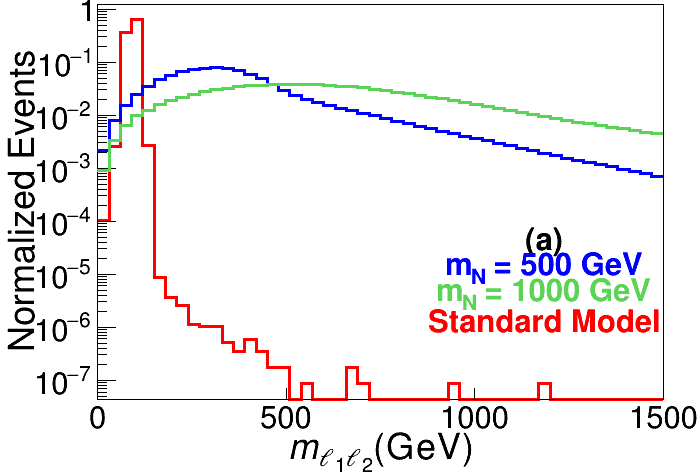}
  \includegraphics[width=0.32\textwidth]{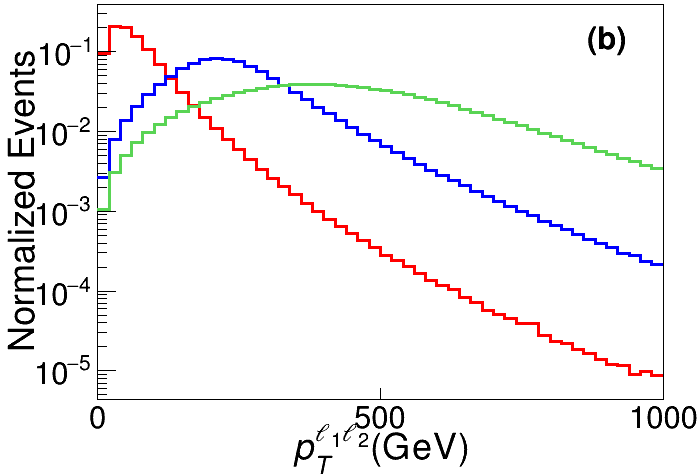}
  \includegraphics[width=0.32\textwidth]{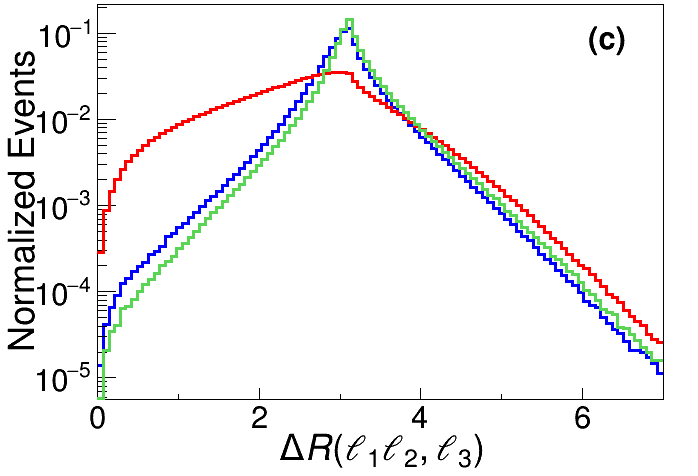}
  \includegraphics[width=0.32\textwidth]{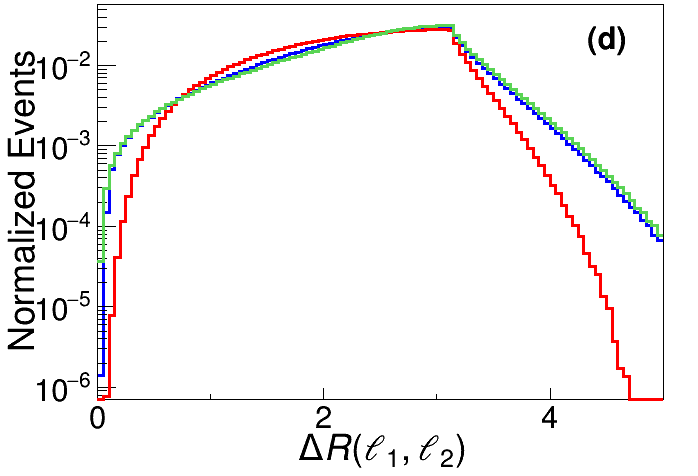}
  \includegraphics[width=0.32\textwidth]{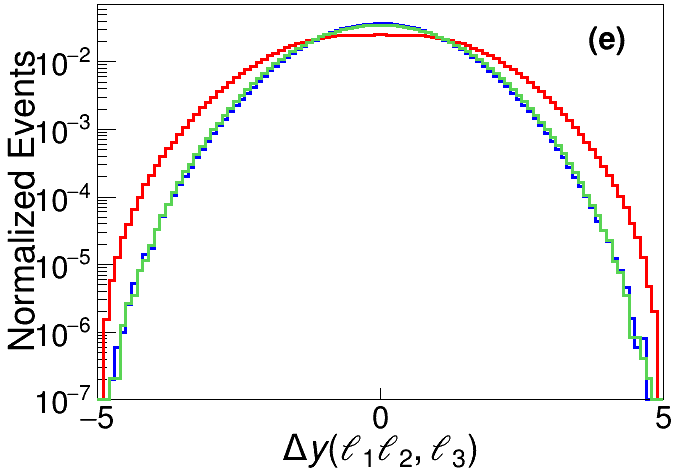}
  \includegraphics[width=0.32\textwidth]{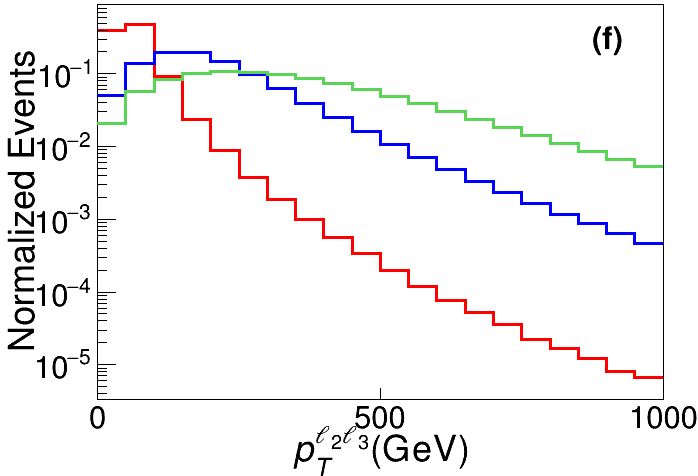}
  \caption{Distributions of background veto observables $m_{\ell_1\ell_2}$ (upper left), $p_T^{\ell_1\ell_2}$ (upper middle), $\Delta R (\ell_1\ell_2, \, \ell_3)$ (upper right), $\Delta R (\ell_1,\, \ell_2)$ (lower left), $\Delta y (\ell_1\ell_2, \, \ell_3)$ (lower middle) and $p_T^{\ell_2\ell_3}$ (lower right) of SM backgrounds (red) and heavy neutrino signals with $m_N = 500$ GeV (blue) and 1000 GeV (green) at $\sqrt{s}=$ 14 TeV. The distributions are normalised so that their integrals are equal to 1.}
  \label{fig:var_bkg}
\end{figure}

In order to further suppress the SM background events, six more feature observables are introduced, and the corresponding distributions are shown in Fig.~\ref{fig:var_bkg}.
\begin{itemize}
\item The invariant mass $m_{\ell_1 \ell_2}$ of the two leptons $\ell_{1}$ and $\ell_{2}$. For background events, $\ell_1$ and $\ell_2$ are from $Z$ boson decay, and $m_{\ell_1 \ell_2}$ forms a peak at the $Z$ mass $m_Z$; for signal events such an observable is rather broad, as seen in the upper left panel of Fig.~\ref{fig:var_bkg}. Therefore, this observable should be the main killer for the background events.
\item The transverse momentum $p_T^{\ell_1 \ell_2}$ of the $\ell_1 \ell_2$ subsystem. For the background events, it is equivalently the transverse momentum $p_T^{Z}$ of the $Z$ boson. As the $W$ and $Z$ boson masses are very close, they fly back-to-back in the background events. In other words, $p_T^W$ and $p_T^Z$ are correlated for background events. In contrast, for the faked $Z$ boson in the signal events, there should be no such a correlation, and the distribution $p_T^{\ell_1 \ell_2}$ is significantly broader, as shown in the upper middle panel of Fig.~\ref{fig:var_bkg}.
\item The angular distance $\Delta R (\ell_1\ell_2, \ell_3)$ between the $\ell_1\ell_2$ subsystem and the charged lepton $\ell_3$ from $W$ boson decay. For the SM backgrounds, when the energies of $W$ and $Z$ bosons are large, they are Lorentz boosted, which can be measured to some extent by the separation $\Delta R (\ell_1\ell_2, \ell_3)$. As seen in the upper right panel of Fig.~\ref{fig:var_bkg}, this separation has a relatively sharper peak for the signal events.
\item The angular distance $\Delta R (\ell_1,\ell_2)$ between the two leptons $\ell_1$ and $\ell_2$ reconstructing the $Z$ boson. For background events, these two leptons are from the $Z$ boson decay. When $Z$ boson is produced near its peak, most of $Z$ bosons have small boost factors. Then the pair of leptons from $Z$ decay are expected to fly back-to-back. The signal events have a relatively broader distribution of $\Delta R (\ell_1, \ell_2)$, as shown in the lower left panel of Fig.~\ref{fig:var_bkg}.
\item The rapidity difference  $\Delta y (\ell_1\ell_2, \ell_3)$ between the $\ell_1\ell_2$ subsystem and the charged lepton $\ell_3$ originating from $W$ boson. As presented in the lower middle panel of Fig.~\ref{fig:var_bkg}, the signal events have a relatively smaller  $|\Delta y (\ell_1\ell_2, \ell_3)|$ than the background events.
\item The transverse momentum $p_T^{\ell_{2}\ell_3}$ of the $\ell_2 \ell_3$ subsystem. As in the signal events these two leptons are from $N$ decay, $p_T^{\ell_{2}\ell_3}$ tends to be larger for the signals than for backgrounds, as seen in the lower right panel of Fig.~\ref{fig:var_bkg}.
\end{itemize}

\begin{figure}[t]
  \centering
  \includegraphics[width=0.45\textwidth]{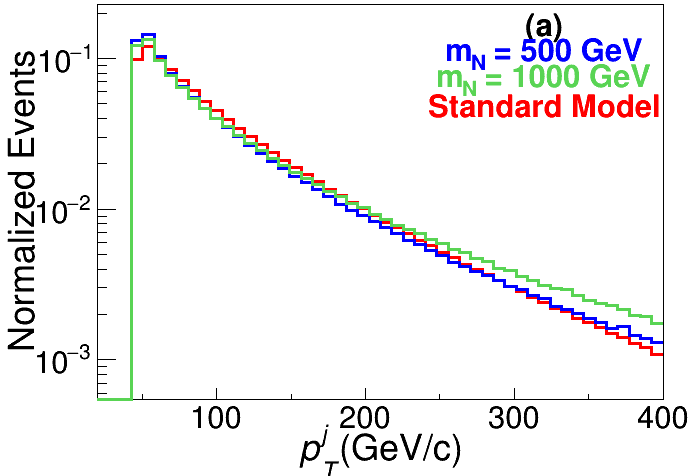}
  \includegraphics[width=0.45\textwidth]{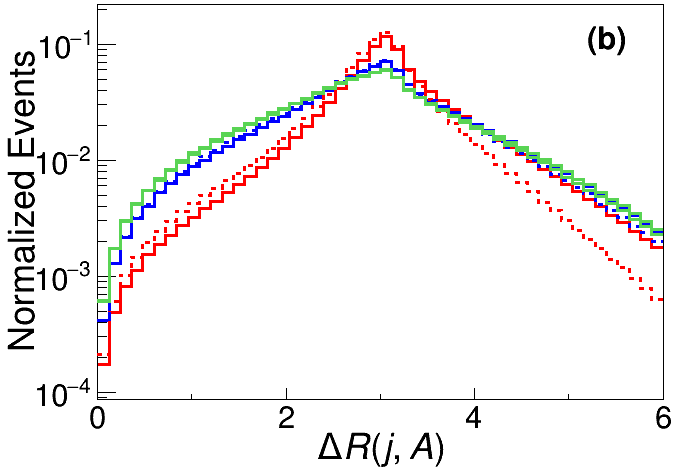}
  \caption{Distributions of jet relevant observables $p_T^j$ (left) and $\Delta R (j,\,A)$ (right) of SM backgrounds (red) and heavy neutrino signals with $m_N = 500$ GeV (blue) and 1000 GeV (green) at $\sqrt{s}=$ 14 TeV. As the variable $p_z^\nu$ is two-folded, the solid lines in the right panel correspond to the solution 0 (larger absolute value of $p_z^{\nu}$) and the dash lines to the solution 1 (smaller absolute value of $p_z^{\nu}$). The distributions are normalised so that their integrals are equal to 1.} 
  \label{fig:var_jet}
\end{figure}

For the pre-selected events with one jet, i.e. the 1-jet channel, three more feature observables are introduced, and their corresponding distributions are shown in Fig.~\ref{fig:var_jet}.
\begin{itemize}
\item The transverse momentum $p_T^{j}$ of jet. The corresponding distributions for the SM backgrounds and signals are shown in the left panel of Fig.~\ref{fig:var_jet}. 
\item The angular distance $\Delta R(j,A)$ between the jet and the rest of the products, i.e. the three charged leptons and neutrino.
As the longitudinal momentum of neutrino has a two-fold ambiguity, $\Delta R(j,A)$ is also two-folded, as seen in the right panel of Fig.~\ref{fig:var_jet}.
\end{itemize}

     
     \begin{figure}[t]
     \centering
     \includegraphics[width=0.75\textwidth]{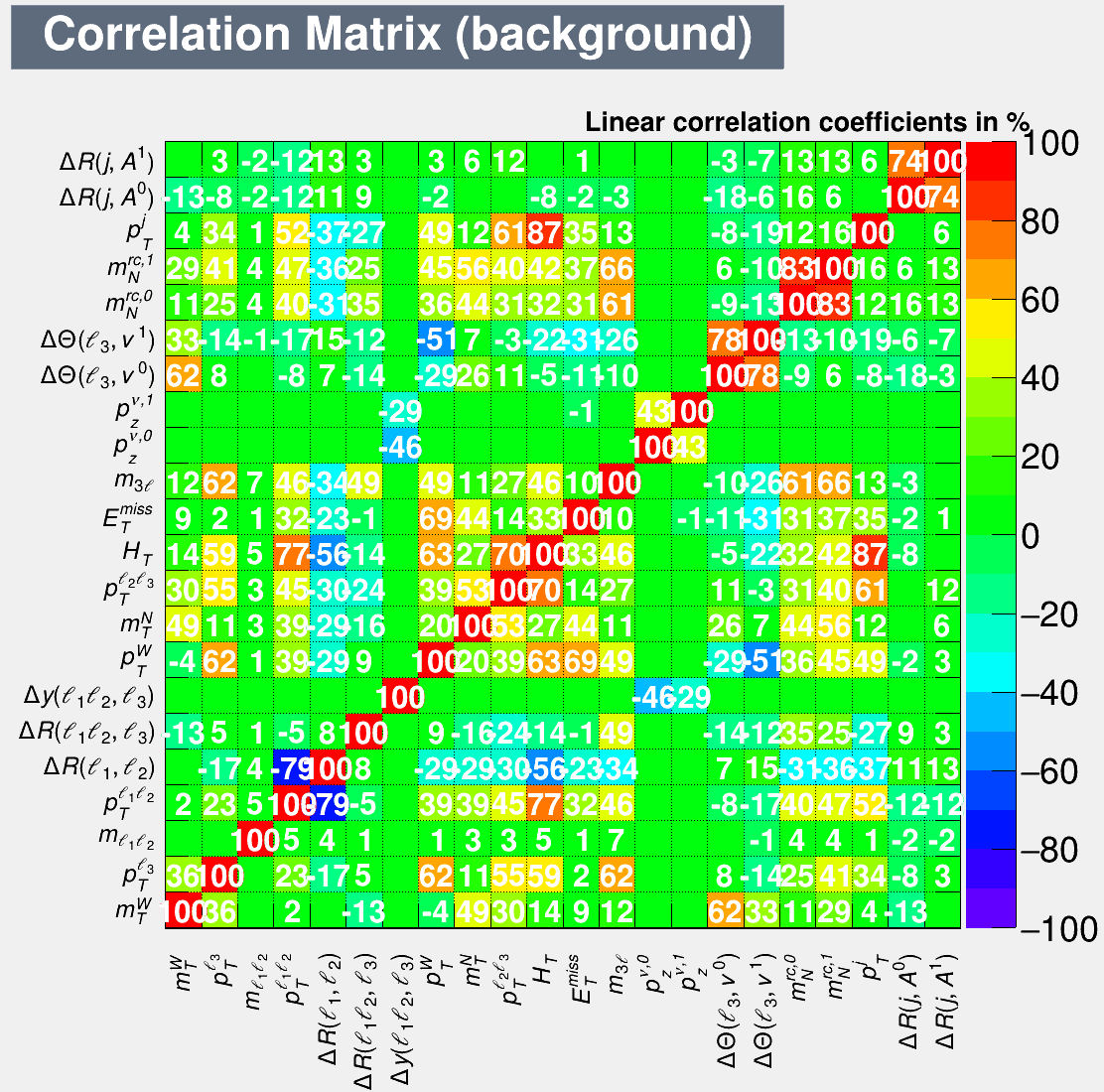}
      \caption{Linear correlation coefficient matrix for backgrounds at $\sqrt{s}=14$ TeV. }
     \label{fig:correlationB}
     \end{figure}
     
     \begin{figure}[t]
     \centering
     \includegraphics[width=0.75\textwidth]{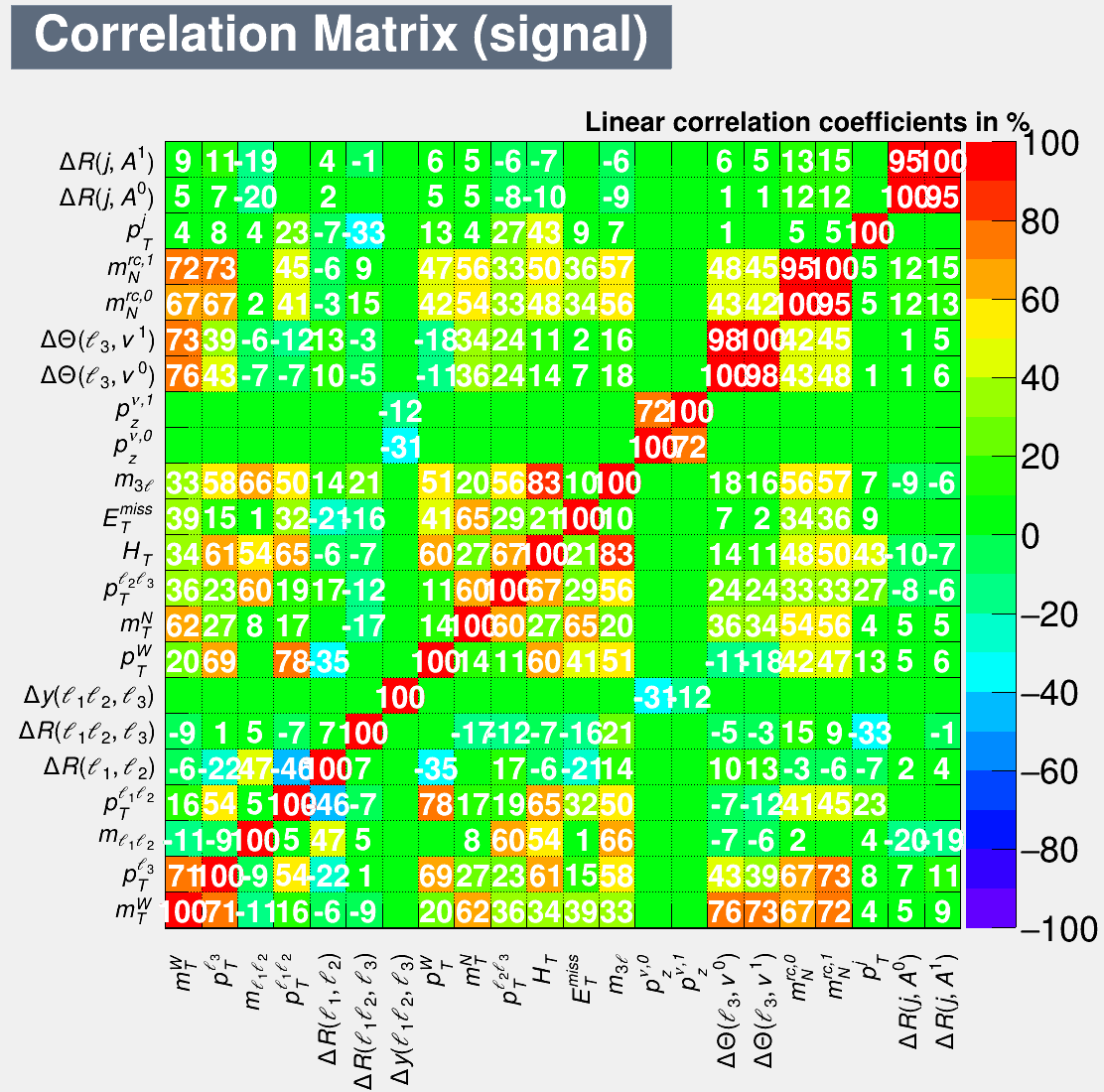}
      \caption{Linear correlation coefficient matrix for the signal with $m_N=500$ GeV at $\sqrt{s}=14$ TeV. }
     \label{fig:correlationS}
     \end{figure}
     
The correlations among the variables above can be quantified by their linear correlation coefficients, which are shown in Figs.~\ref{fig:correlationB} and \ref{fig:correlationS}, respectively, for backgrounds and the signal with $m_N = 500$ GeV at $\sqrt{s}=14$ TeV. It turns out that most of the coefficients are close to zero. Variable decorrelation via the square-root of the covariance matrix is the first step of the TMVA training processes, and ML methods take the correlation matrices into account during the training processes.

It is very clear in Fig.~\ref{fig:var_jet} that the distributions of $p_T^{j}$ and $\Delta R(j,A)$ are rather similar for the SM backgrounds and signal. In other words, the ML analysis in Section~\ref{sect:ML} will be dominated by the events without any jet. Since events with jets contribute little to the discrimination of signal from backgrounds in our analysis and the simulations with two (or more) jets cost a lot of computing resources, we will not consider the channels with two (or more) jets from now on. 
Here we describe the procedure how we perform our analysis. 
\begin{enumerate}
    \item In the MC generation, we produce 0-jet and 1-jet events together. In the multi-variate analyses, we divide our data sample into two separate subgroups, i.e. 0-jet and 1-jet data samples. 
    \item For each heavy neutino mass, we perform ML training for the 0-jet and 1-jet groups, respectively, and obtain the best performance to separate signal and background events. 
    \item Finally, we combine these two subgroups of data samples to get the prospects of $|V_{\ell N}|^2$ at the high-energy colliders.
\end{enumerate}




\begin{figure}[t]
  \centering
  \includegraphics[width=0.49\textwidth]{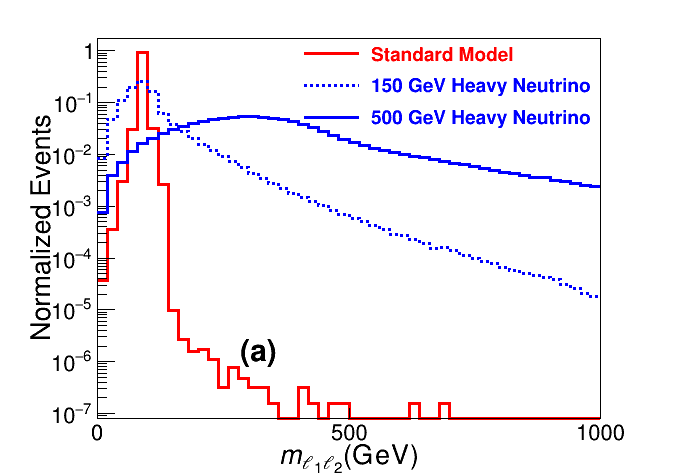}
  \includegraphics[width=0.49\textwidth]{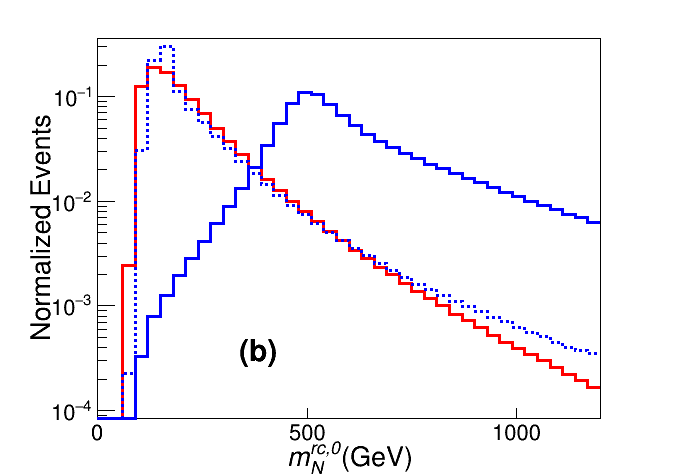}
  \caption{Distributions of $m_{\ell_1\ell_2}$ (left) and $m_N^{\rm rc,0}$ (right) of SM backgrounds (red) and heavy neutrino signals with $m_N = 150$ GeV (dashed blue) and 500 GeV (solid blue) at $\sqrt{s}=$ 14 TeV. The distributions are normalised so that their integrals are equal to 1.}
  \label{fig:150_500}
\end{figure}


It is noteworthy that the separation power of the variables above varies for different beam energies and heavy neutrino masses.  For example, for most center-of-mass energies and heavy neutrino mass $m_N$, it is found that the most powerful observable is $m_{\ell_1\ell_2}$. As shown in the upper left panel of Fig.~\ref{fig:var_bkg}, the SM background events form a sharp peak near the narrow window at the $Z$ mass of 91 GeV, while for the signal events, the peak value of $m_{\ell_1 \ell_2}$ moves in term of  the heavy neutrino mass. 
It is remarkable in the left panel of Fig.~\ref{fig:150_500} that when $m_N$ is around 150 GeV, the observable $m_{\ell_1\ell_2}$ for the signal peaks at around 90 GeV, which makes the separation of signal and background events rather challenging. It is found that other observables have also less separation power for the case of $m_N=150$ GeV. As an explicit example, the distributions of $m_N^{\rm rc,0}$ for backgrounds and the signal cases $m_N=150$ GeV and 500 GeV are shown in the right panel of Fig~\ref{fig:150_500}, which are denoted, respectively, by the red, dashed blue and solid blue lines. It is clear that for the case of $m_N=150$ GeV, the distribution of $m_N^{\rm rc,0}$ for signal almost overlaps with that for the SM backgrounds.

\subsection{Machine learning methods and analysis}
\label{sect:ML}

\begin{figure}[t]
  \centering
  \includegraphics[width=0.49\textwidth]{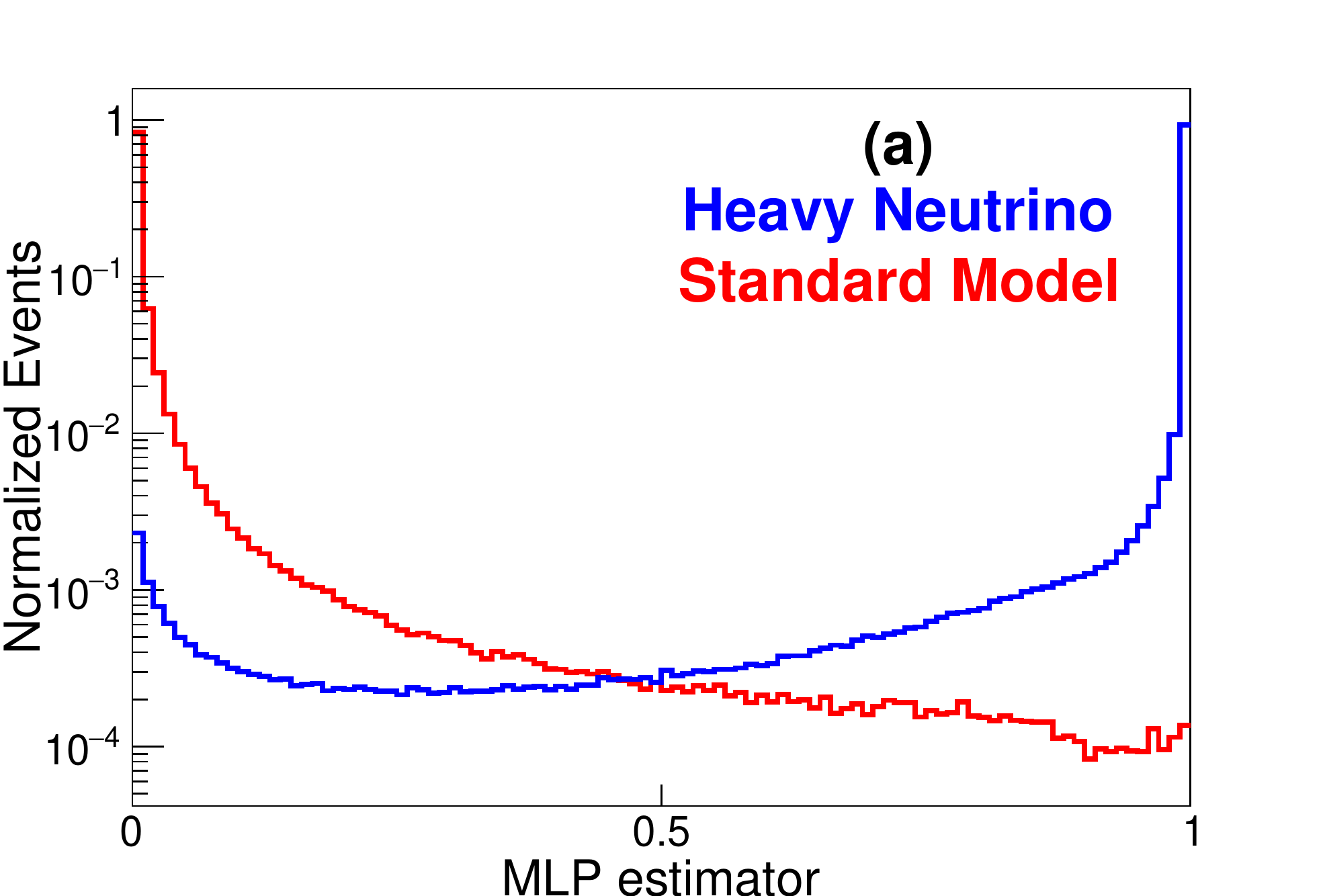}
  \includegraphics[width=0.49\textwidth]{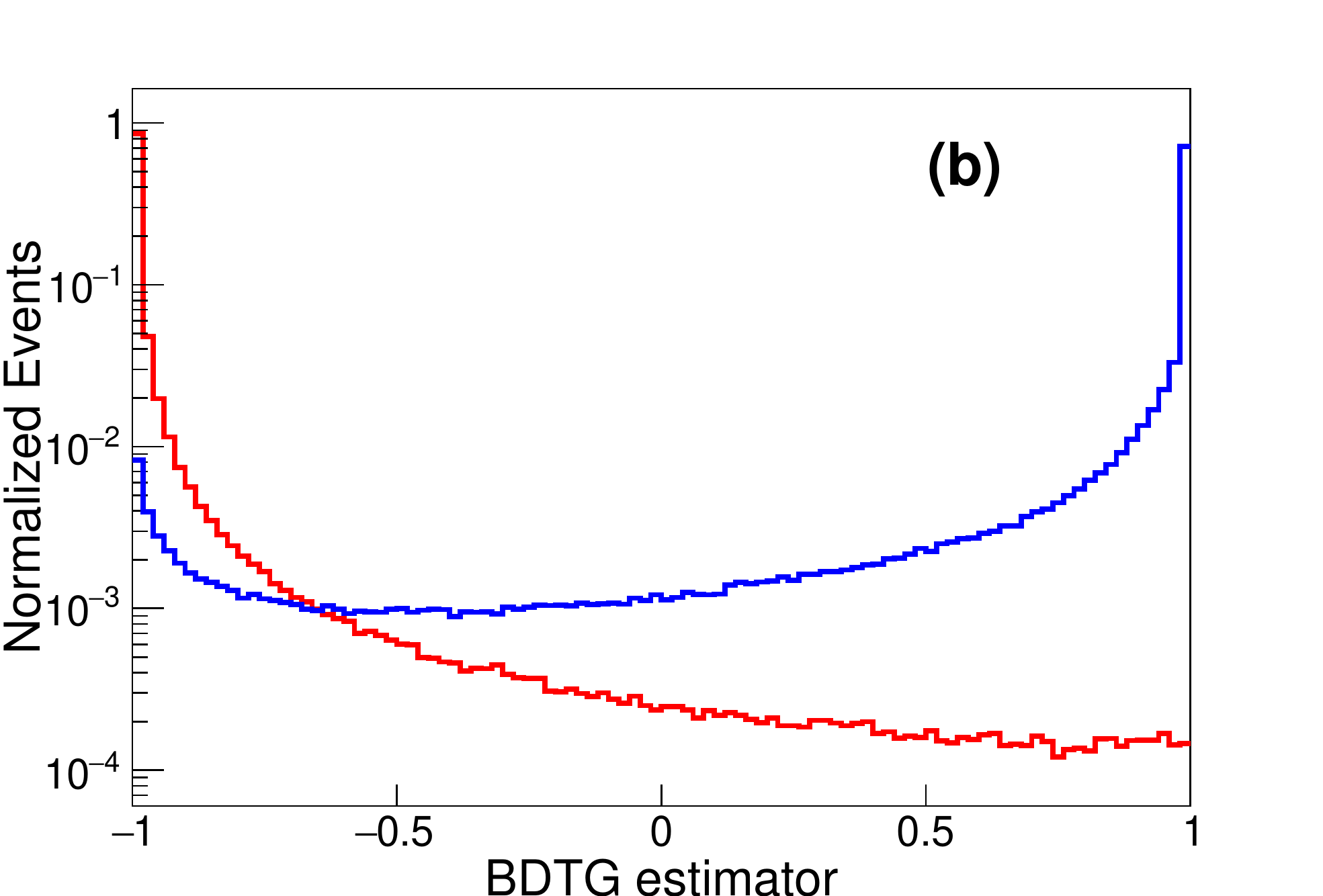}
  \caption{ML estimator distributions of SM backgrounds (red) and heavy neutrino signal (blue) with $m_N = 500$ GeV at $\sqrt{s}=$ 14 TeV. The distributions are normalised so that their integrals are equal to 1. The left and right panels are for the MLP and BDTG methods, respectively.  }
  \label{fig:estimator}
\end{figure}

As mentioned in the introduction, we will use ML to achieve the optimal separation power between the background and signal events. In particular, we adopt two multi-variate analysis methods, i.e.  the MLP~\cite{Fukushima:2013b} and BDTG~\cite{10.2307/2699986} techniques, which have been encoded in the TMVA package~\cite{2007physics...3039H}. 
A portion of MC samples are used for ML training. Input variables are prepossessed for decorrelation via the square-root of the covariance matrix (cf. Figs.~\ref{fig:correlationB} and \ref{fig:correlationS}). Nodes are constructed in the ML algorithms. The number of nodes is related to the number of trees in BDTG, while that is related to the number of hidden layers and the number of nodes on each layer in MLP. There are weights connecting those nodes.  In ML methods, more useful variables have larger weights such as $m_{\ell_1 \ell_2}$ and $m_N^{\rm rc}$, while less useful variables have small weights, for instance $p_T^j$, $\Delta R(j,A)$, $\Delta R(\ell_1, \ell_2)$ and $\Delta y (\ell_1\ell_2,\ell_3)$. 
A loss function is defined to describe the sum of difference between the machine prediction (the probability to be a signal event) and data (1 for signal and 0 for background). The goal of training process is to adjust the weights to minimize the loss function. Since the weights are free parameters in the ML, the number of weights can not be too large compared to the sizes of the samples, otherwise the ML will be over-fitted. In our case, the number of trees is set to be 1000 in BDTG. In MLP there is one layer, where there are 25 nodes. 
    
These two methods combine the multi-dimensional feature observables in Figs.~\ref{fig:var_global} to \ref{fig:var_jet} into a single likelihood variable. The 0-jet and 1-jet events above are combined based on their production cross sections (cf. Fig.~\ref{fig:xs}). However, the jet relevant distributions in Fig.~\ref{fig:var_jet} do not contribute much to the separation power in the ML analysis, as compared to the distributions of charged leptons and MET in Figs.~\ref{fig:var_global} to \ref{fig:var_bkg}. 
The distributions of one-dimensional likelihood variables in MLP and BDTG for the backgrounds and signal with $m_N = 500$ GeV at $\sqrt{s}=14$ TeV are shown in the left and right panels of Fig.~\ref{fig:estimator}, respectively. The background and signal distributions are denoted, respectively, by the red and blue lines. It is clear that both the two methods can well distinguish the signal from backgrounds.

\begin{figure}[t]
  \centering
  \includegraphics[width=0.49\textwidth]{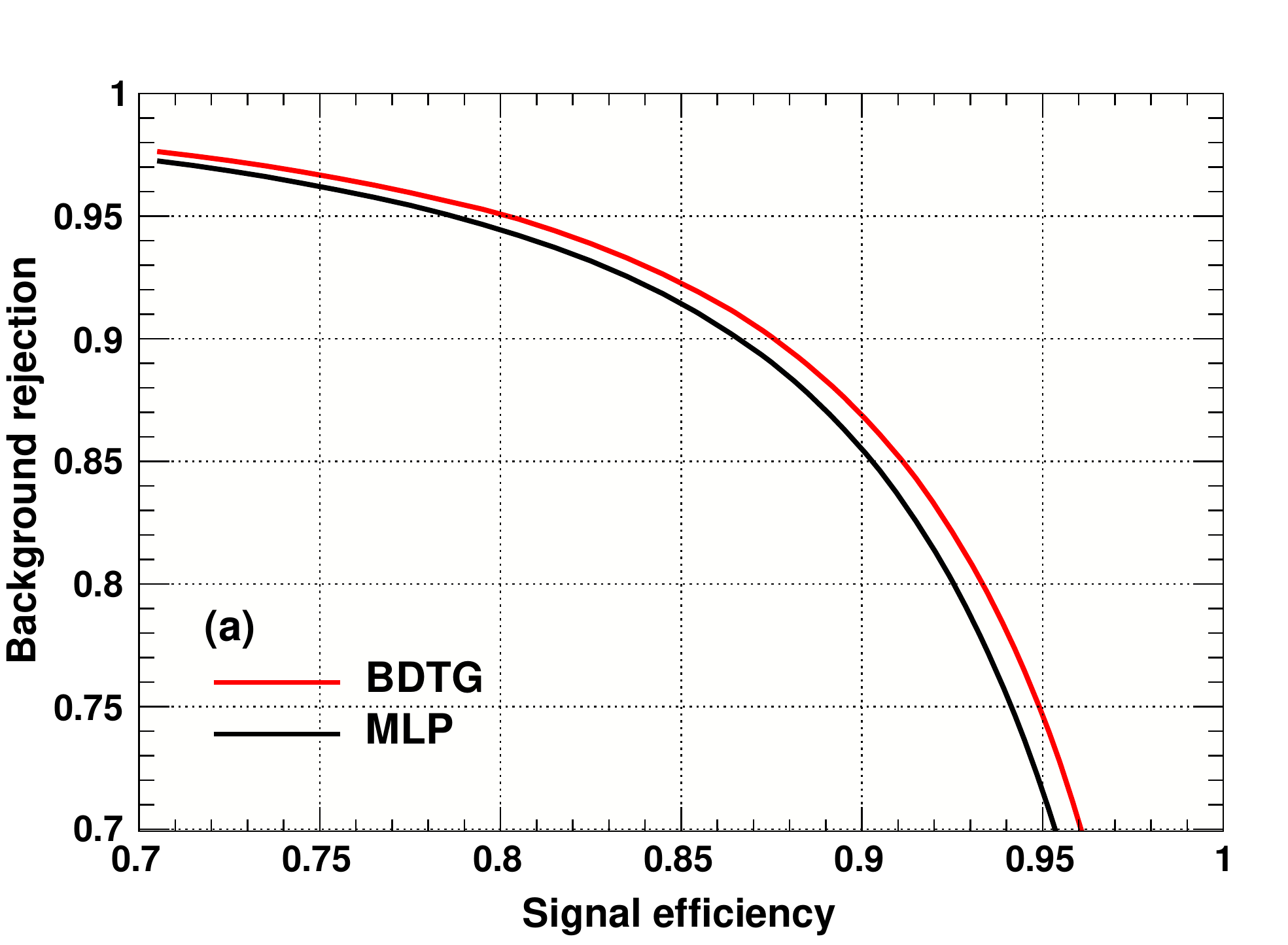}
  \includegraphics[width=0.49\textwidth]{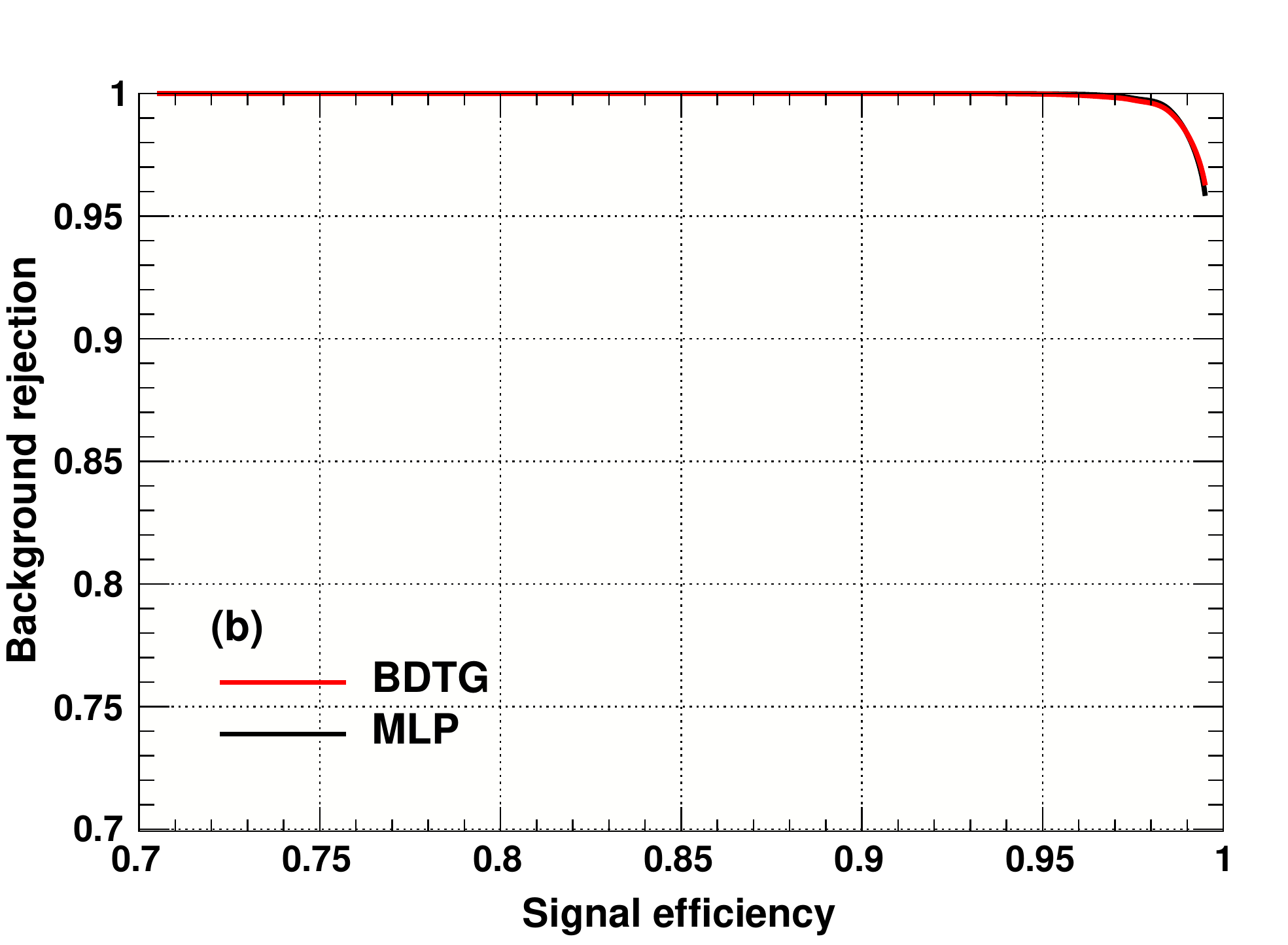}
  \caption{ Background rejection versus signal efficiency from BDTG (red) and MLP (black) at $\sqrt{s}=$ 14 TeV, for the scenarios $m_N = 150$ GeV (left) and 500 GeV (right).   }
  \label{fig:ROC}
\end{figure}

After optimizing the ML estimator cuts, we can achieve good background rejections (i.e. the portion of rejected background w.r.t. the total) and high efficiencies to pick out signal events (i.e. the portion of signal left w.r.t. the total). As an explicit example, the rejection curves for the two methods BDTG and MLP with $m_N = 500$ GeV are demonstrated in the right panel of Fig.~\ref{fig:ROC}, as functions of the signal efficiency. As a comparison, the rejection curves for the case $m_N = 150$ GeV are shown in the left panel of Fig.~\ref{fig:ROC}. In both the two panels, the BDTG and MLP methods are depicted in red and black, respectively. Comparing the two panels of Fig.~\ref{fig:ROC}, the separation performance of the ML estimators is significantly better at $m_N=500$ GeV than that at 150 GeV (cf.  Fig.~\ref{fig:150_500}). For both the two heavy neutrino scenarios in Fig.~\ref{fig:ROC}, the BDTG analysis gives slightly better discrimination performance than MLP, especially at high signal efficiency. In general, MLP and BDTG methods yield compatible results. When the number of input observables is the same, the BDTG method can be about 7.5 times faster than the MLP method in training processes. Therefore, in this work we will use the results of the BDTG method and take the MLP method as a cross check.

\begin{figure}[t]
  \centering
  \includegraphics[width=0.49\textwidth]{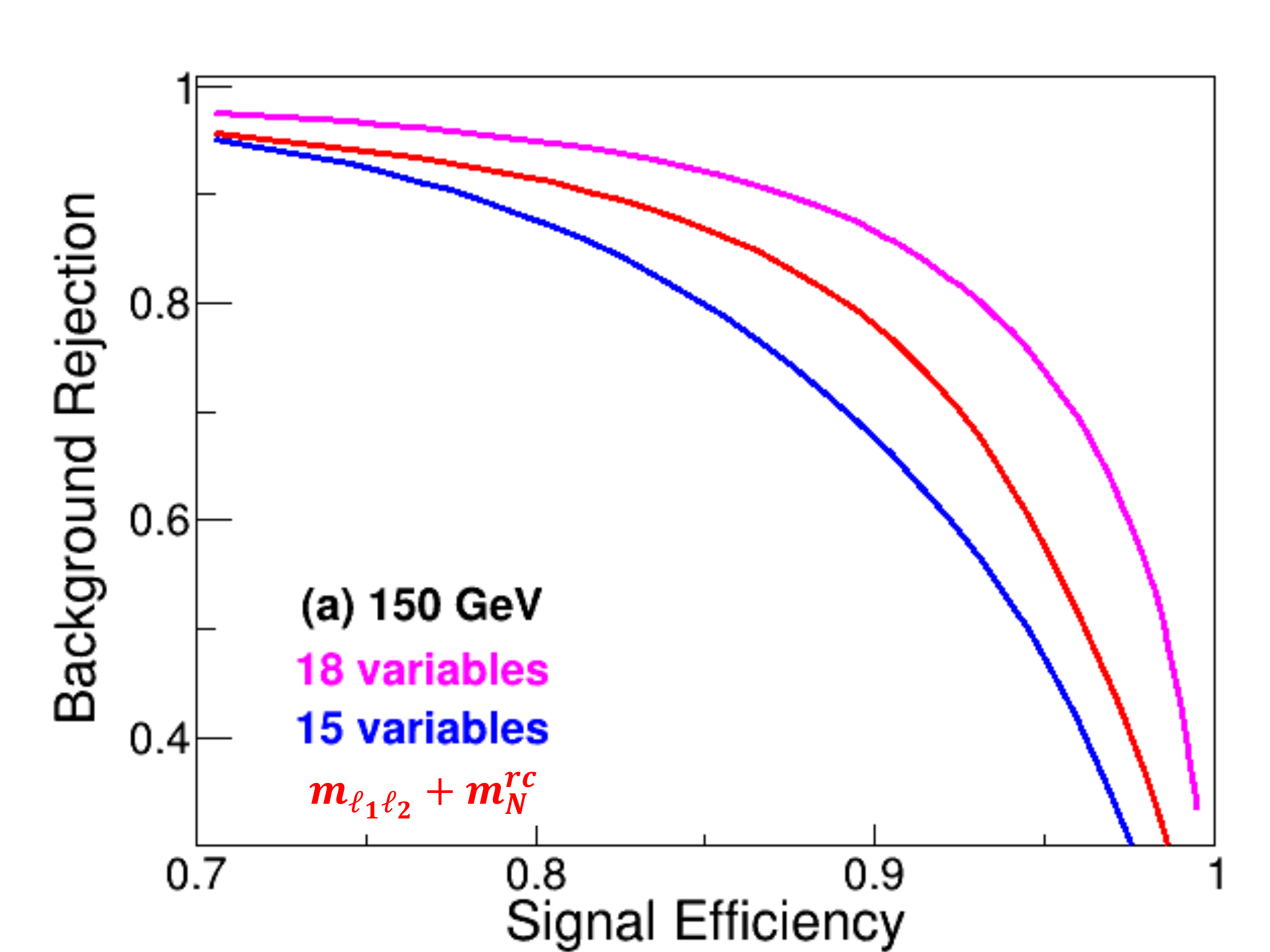}
  \includegraphics[width=0.49\textwidth]{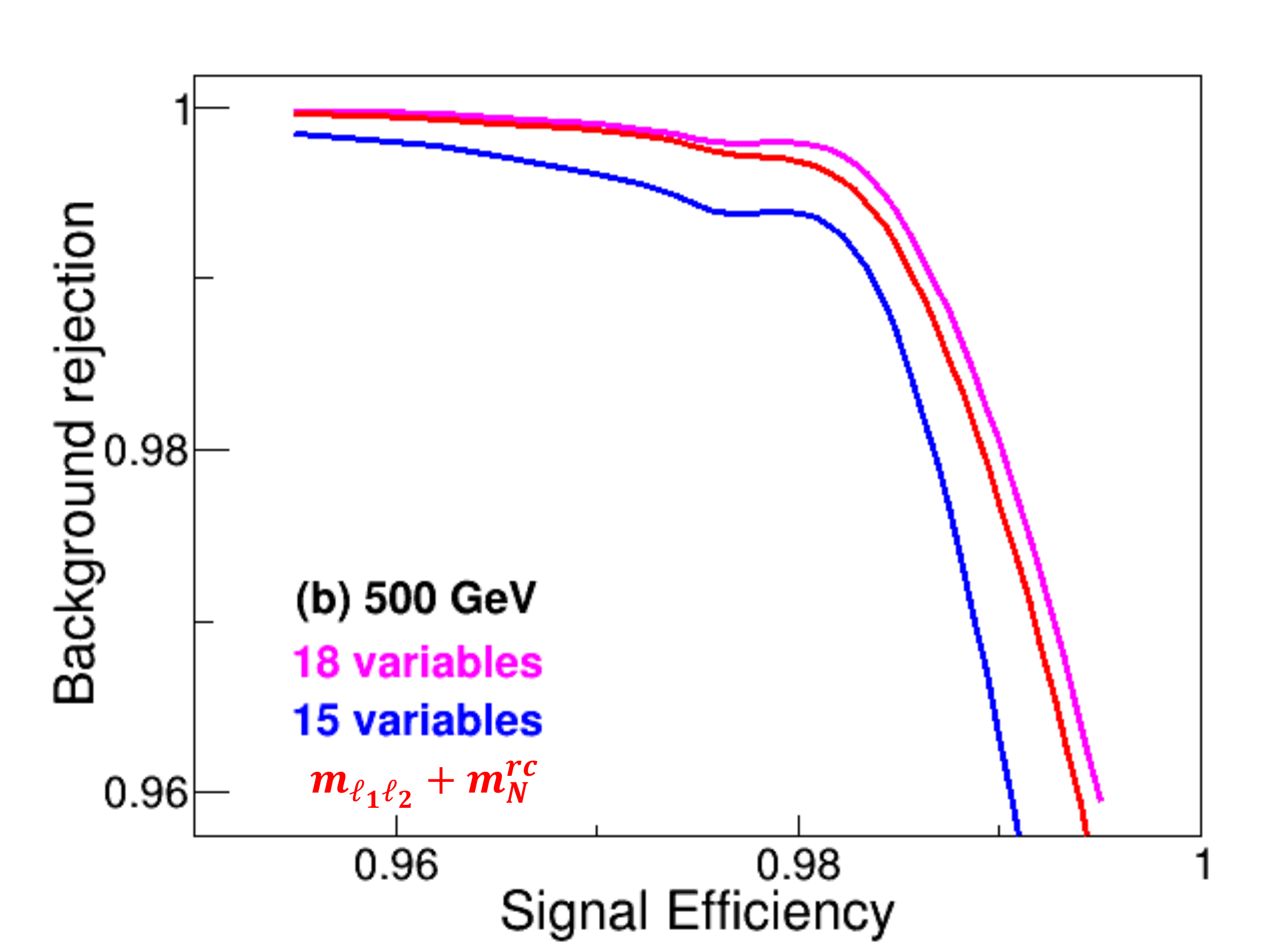}
  \caption{Background rejection versus signal efficiency for different sets of variables in MLP at $\sqrt{s}=$ 14 TeV in the 0-jet channel for $m_N = 150$ GeV (left) and 500 GeV (right). The magenta, red and blue lines are, respectively, for all the 18 variables in Fig.~\ref{fig:var_global} to \ref{fig:var_bkg}, the variables $m_{\ell_1 \ell_2}$, $m_N^{\rm rc,0}$ and $m_N^{\rm rc,1}$, and the rest 15 variables. }
  \label{fig:150_500_ZN}
\end{figure}

It is remarkable in Figs.~\ref{fig:var_global} to \ref{fig:var_bkg} that the most distinguishing feature observables are $m_{\ell_1 \ell_2}$ and $m_N^{\rm rc}$. To show the separation power of these observables, the 
rejection curves for the observables $m_{\ell_1 \ell_2}$ and the $m_N^{\rm rc}$, the other 15 observables in Figs.~\ref{fig:var_global} to \ref{fig:var_bkg} and all the 18 observables are shown in Fig.~\ref{fig:150_500_ZN}, respectively, as the red, blue and magenta lines. The left and right panels are for the cases of $m_N = 150$ GeV and 500 GeV, respectively. 
As clearly seen in Fig.~\ref{fig:150_500_ZN}, the discrimination power is dominated by $m_{\ell_1 \ell_2}$ and $m_N^{\rm rc}$ for both the two scenarios. For the challenging case $m_N=150$ GeV, including the other 15 observables can undoubtedly improve the yield, where the distributions of $m_{\ell_1 \ell_2}$ and $m_N^{\rm rc}$ overlap for the signal and backgrounds (cf. Fig.~\ref{fig:150_500}). 

Compared to the cut-based analysis, the ML methods can not only use the full information of the distributions in Figs.~\ref{fig:var_global} to \ref{fig:var_jet}, but also take into account the correlations among the variables (cf.  Figs.~\ref{fig:correlationB} and \ref{fig:correlationS}). Optimizing the discrimination power of the signal and background data sets, the ML analysis can significantly improve the prospects of $|V_{\ell N}|^2$ at the future high-energy hadron colliders. As seen in Fig.~\ref{fig:Bln} below, the mixing angles $|V_{e N}|^2$ and $|V_{\mu N}|^2$ can be probed down to ${\cal O}(10^{-6})$ to ${\cal O}(10^{-5})$ for $m_N =100$ GeV at the future hadron colliders, which is roughly one order of magnitude better than the current sensitivities in the literature.

\section{Sensitivities of $|V_{\ell N}|^2$ at future hadron colliders}
\label{sec:sensitivity}

The prospects of the heavy-light neutrino mixing parameter $|V_{\ell N}|^2$ at the high-energy $pp$ colliders depend on the heavy neutrino mass $m_N$ and the cut value of ML estimators as well as the colliding energy and total luminosity. For a given collision energy and heavy neutrino mass $m_N$, after setting the ML estimator cuts, 
the number of background ($B$) and signal ($S$) events and the resultant significance $S/\sqrt{S+B}$ can be computed. A genetic algorithm {\tt GATOR}~\cite{Luo:2019qbk} is used to scan the cut values of the ML estimators 
to find the $95\%$ C.L. sensitivities of $|V_{\ell N}|^2$ with $S/\sqrt{S+B} = 2$.
We have chosen three typical collision energies and integrated luminosities, i.e. the HL-LHC 14 TeV with an integrated luminosity of 3 ab$^{-1}$, the HE-LHC with $\sqrt{s}=27$ TeV and 15 ab$^{-1}$, and the future 100 TeV collider with  30 ab$^{-1}$.

\begin{figure}[t]
  \centering
  \includegraphics[width=0.65\textwidth]{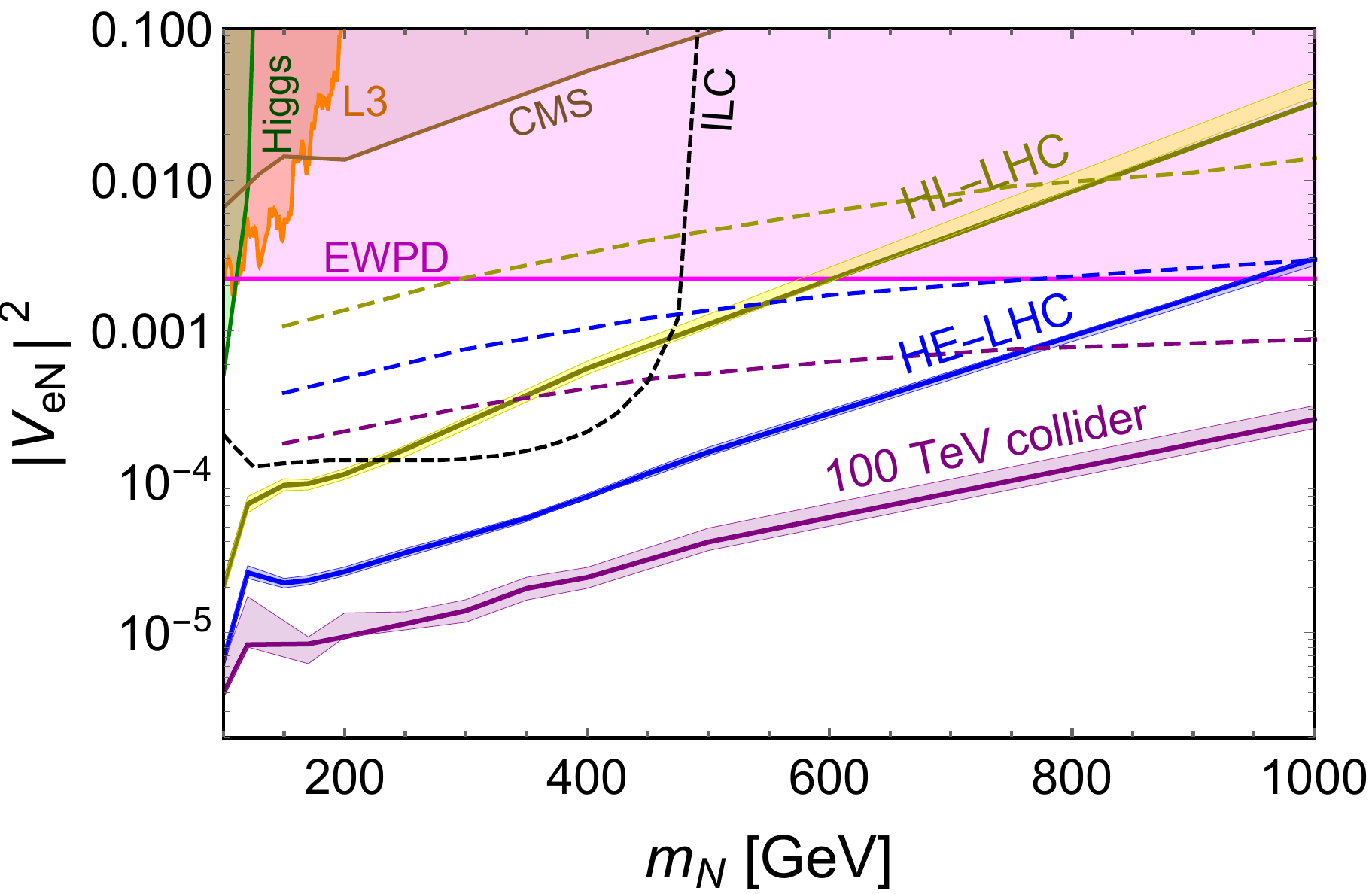}
  \includegraphics[width=0.65\textwidth]{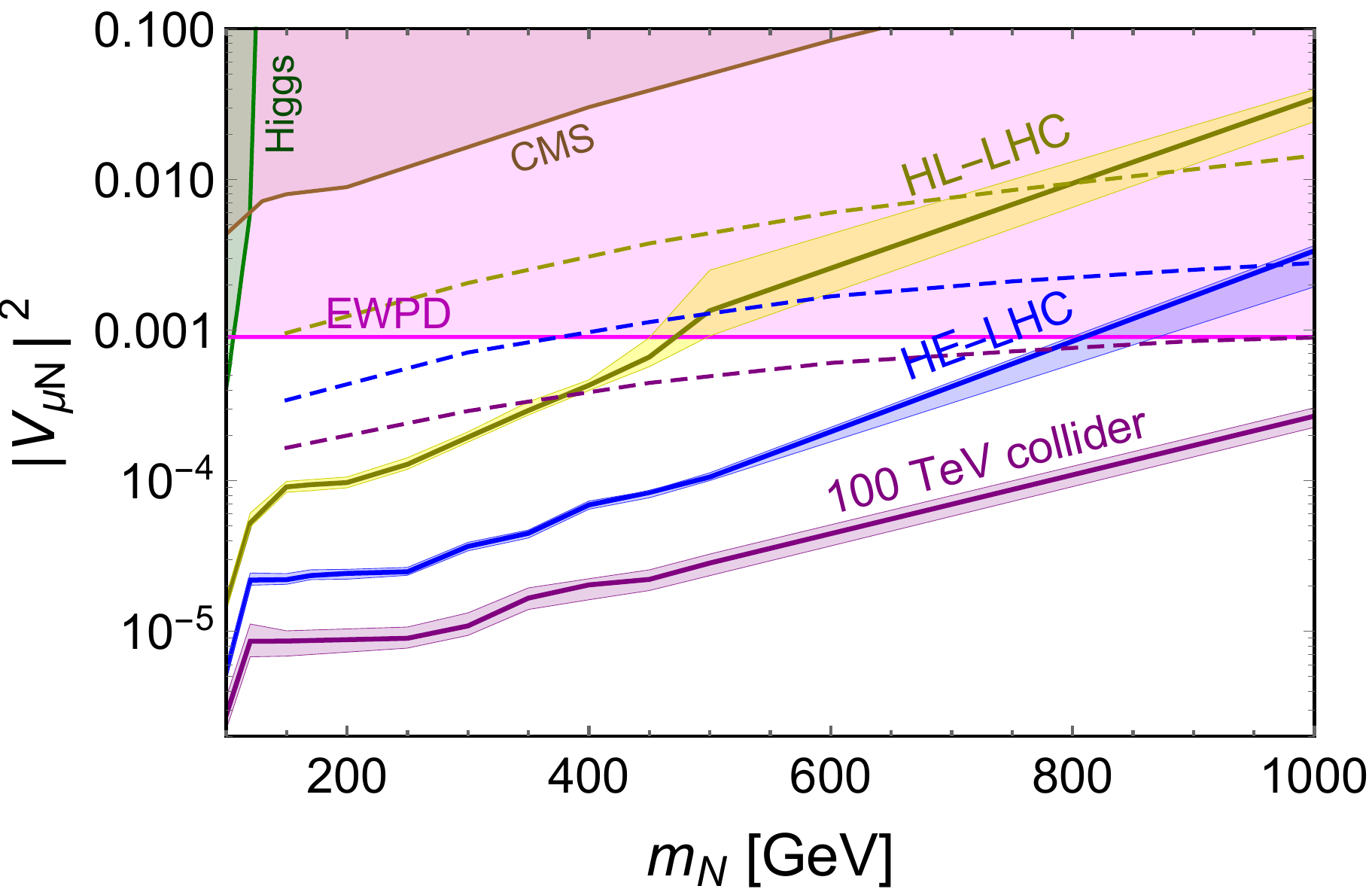}
  \caption{Sensitivities of the heavy-light neutrino mixing $|V_{e N}|^2$ (upper) and $|V_{\mu N}|^2$ (lower) at 95\% C.L. as function of heavy neutrino mass $m_N$ from the ML analysis of the trilepton signal in Eq.~(\ref{eqn:trilepton}). 
  The solid darker yellow, blue and purple lines correspond, respectively, to the central values of sensitivities at the HL-LHC 14 TeV with a luminosity of 3 ab$^{-1}$, HE-LHC 27 TeV with 15 ab$^{-1}$ and the 100 TeV collider with 30 ab$^{-1}$. The lighter color bands are due to the systematic uncertainties. The dashed lines are the corresponding sensitivities using dynamic jet vetoes from Ref.~\cite{Pascoli:2018heg}. The dashed black line in the upper panel is the sensitivity at the ILC 500 GeV with 500 fb$^{-1}$. The shaded regions are excluded by the direct searches of $N$ by L3 (orange)~\cite{L3:1999ymc, L3:2001zfe}, CMS (brown)~\cite{CMS:2018iaf}, precision SM Higgs data (green)~\cite{Das:2017zjc}, and the indirect limits from EWPD (magenta)~\cite{delAguila:2008pw, Akhmedov:2013hec, Antusch:2014woa, Blennow:2016jkn}. See text for more details.  }
  \label{fig:Bln}
\end{figure}

The prospects of $|V_{eN}|^2$ and $|V_{\mu N}|^2$ at the high-energy hadron colliders are shown in the upper and lower panels of Fig.~\ref{fig:Bln}, respectively, as function of the heavy neutrino mass $m_N$. 
Next-to-leading order corrections have been taken into consideration: the cross section of signal is enlarged by a factor of 16\%~\cite{Das:2016hof, Degrande:2016aje, Pascoli:2018heg}, while that for the SM backgrounds by a factor of 30\%, including the contributions from the additional off-shell and $4\ell$ processes~\cite{PhysRevD.50.1931, Campanario:2010hp, Pascoli:2018heg}.
The corresponding sensitivities of $|V_{\ell N}|^2$ at the HL-LHC, HE-LHC and future 100 TeV collider are shown as the darker yellow, blue and purple lines in Fig.~\ref{fig:Bln}, respectively. 
We have also estimated the systematic errors, which include the uncertainties in the cross sections due to MC statistics, scale variation and the PDF variation for both backgrounds and signal. It is found that the systematic error associated with the MC statistics is less than 5\% over the entire heavy neutrino mass range in this paper. The uncertainties in the cross sections due to scale variation and PDF variation are at the order of 10\% and $2\%$, respectively. The resultant systematic uncertainties of $|V_{\ell N}|^{2}$ are $\lesssim$ 10\%. The effects of systematic uncertainties on the sensitivities of $|V_{\ell N}|^2$ are indicated by the lighter color bands in Fig.~\ref{fig:Bln}. As implied by the distributions in Fig.~\ref{fig:150_500}, the sensitivities at $m_N \sim 150$ GeV are relatively weak. 

There are a large variety of laboratory, astrophysical and cosmological limits on the mixing of heavy neutrinos with the active neutrinos. See Refs.~\cite{Deppisch:2015qwa,Bolton:2019pcu} for a complete list for current experimental limits on $|V_{eN}|$ and $|V_{\mu N}|$. When the heavy neutrino mass is larger than roughly 100 GeV, the constraints are mostly from the direct searches of $N$ at the LEP and LHC, the precision Higgs data and EWPD.
\begin{itemize}
\item The heavy neutrinos have been searched for at the LEP by the L3 collaboration via the process $e^+ e^- \to N_\alpha \nu_\alpha$~\cite{L3:1999ymc, L3:2001zfe}, with heavy neutrino mass up to 205 GeV. For a 100 GeV $N$, the mixing $|V_{eN}|^2$ can be constrained to the order of 0.002, and is shown as the orange shaded region in the upper panel of Fig.~\ref{fig:Bln}. There are also searches of heavy neutrinos from $Z$ boson decay via $e^+ e^- \to Z \to \nu N$ for $m_N$ below the $Z$ boson mass by L3~\cite{L3:1990aqk, L3:1992xaz}, DELPHI~\cite{DELPHI:1991yjf, DELPHI:1996qcc} and OPAL~\cite{OPAL:1990fcc}. These limits are not shown in Fig.~\ref{fig:Bln}.

\item 
The searches of heavy Majorana neutrinos in the LNV signals have been performed by ATLAS~\cite{ATLAS:2015gtp, ATLAS:2019kpx} and CMS~\cite{CMS:2012wqj, CMS:2015qur, CMS:2016aro, CMS:2018szz, CMS:2018jxx}. However, the LHC searches of LNV signals are not applicable to the Dirac neutrinos in this paper. To set limits on the Dirac $N$, we adopt the limits from the CMS data~\cite{CMS:2018iaf}, which utilize all the possible combinations of electrons and muons in the process $pp \to W^{(\ast)} \to \ell\ell\ell \nu$ including both LNV and lepton number conserving signals from Majorana neutrinos. In absence of signal, these data can be interpreted as conservative limits on a Dirac neutrino $N$. The CMS limits on $m_N$ and $|V_{\ell N}|^2$ (with $\ell = e,\, \mu$) are presented as the brown shaded regions in Fig.~\ref{fig:Bln}. 

\item In the inverse seesaw, the heavy neutrino couples to the SM Higgs doublet via the first term in Eq.~(\ref{eqn:Lagrangian}). When the heavy neutrino is lighter than the SM Higgs boson, it can be produced from SM Higgs via the exotic decay $h \to \nu N \to \ell\ell\nu\nu$~\cite{Das:2017zjc}, which has the same final states as the $h \to WW^\ast$ channel. Therefore the neutrino mixing angle $|V_{\ell N}|$ can be constrained by the Higgs $WW^\ast$ data. Adopting the LHC 8 TeV data~\cite{ATLAS:2014aga}, the limits on $m_N$ and $|V_{\ell N}|$ in Ref.~\cite{Das:2017zjc} are shown as the green shaded regions in Fig.~\ref{fig:Bln}, which exclude the heavy-light neutrino mixing $|V_{\ell N}|^2$ up to ${\cal O}(10^{-3})$ for a 100 GeV heavy neutrino. 

\item Through the heavy-light neutrino mixing $V_{\ell N}$, the heavy neutrino couples to the $W$ boson, which induces the double-beta collider process $W^\pm W^\pm \to \ell^\pm \ell^\pm$ if $N$ is a Majorana fermion~\cite{Fuks:2020att}. In analogous to the case of neutrinoless double-beta decays, the heavy neutrino plays the role of mediator in the double-beta collider process, and can not be produced directly at colliders.  It is found that at HL-LHC 13 TeV with a luminosity of 3 ab$^{-1}$, the mixing angle $|V_{\mu N}|^2$ can be probed down to 0.03~\cite{Fuks:2020att}. However, such a process is not relevant to the Dirac heavy neutrino in this paper.

\item The heavy-light neutrino mixing will induce the non-unitarity of light neutrino mixing matrix $V_{\rm PMNS}$, and thus has effects on the EW precision observables, such as the measurements of weak mixing angle, the $W$ boson mass, the precise $W$ and $Z$ boson decay data,  the lepton universality tests in the meson, tau and $W$ boson decays and the tests of unitarity of CKM matrix~\cite{delAguila:2008pw, Akhmedov:2013hec, Antusch:2014woa, Blennow:2016jkn}. It turns out that the EWPD exclude the mixing angle $|V_{eN}|^2 >2.2\times10^{-3}$ and $|V_{\mu N}|^2 >9.0 \times10^{-4}$, as indicated by the magenta shaded regions in Fig.~\ref{fig:Bln}.
\end{itemize}

There have been intensive studies of the searches of heavy neutrinos at the high-energy lepton colliders. 
For instance, the heavy neutrino can be searched in the channel $e^+ e^- \to \nu N$ at the ILC, which is mediated by a $Z$ boson in the $s$-channel or a $W$ boson in the $t$-channel. With $\sqrt{s}=500$ GeV and an integrated luminosity of 500 fb$^{-1}$, the mixing angle $|V_{eN}|^2$ can be probed up to the order of $10^{-4}$ for $m_N \lesssim 400$ GeV~\cite{Banerjee:2015gca}, as indicated by the dashed black line in the upper panel of Fig.~\ref{fig:Bln}. At the CLIC with $\sqrt{s} = 1.5$ TeV and 3 TeV, the heavy neutrino can be probed to a larger mass, but the decay products from heavy neutrino will be more collimated and form fat jets; more details can be found e.g. in Refs.~\cite{Chakraborty:2018khw, Das:2018usr}. 

The hadron collider searches of heavy neutrinos in the trilepton signal can be found e.g. in Refs.~\cite{Das:2014jxa, Das:2015toa,  Das:2016hof, Antusch:2016ejd, Das:2018hph, Pascoli:2018heg, Pascoli:2018rsg}, 
and it is found that in this channel the relative amounts of leptonic and hadronic activities on an event-by-event basis are very useful to suppress the SM backgrounds~\cite{Pascoli:2018heg}. Considering both the trilepton process in Eq.~(\ref{eqn:trilepton}) and the VBF $W^\ast \gamma$ process which also generates three charged leptons plus MET and jets, the prospects of $|V_{eN}|^2$ and $|V_{\mu N}|^2$ from Ref.~\cite{Pascoli:2018heg} are shown as the dashed lines in the upper and lower panels of Fig.~\ref{fig:Bln}, respectively. The dashed yellow, blue and purple lines are, respectively, for the HL-LHC with $\sqrt{s} = 14$ TeV and the luminosity of 3 ab$^{-1}$, HE-LHC 27 TeV with 15 ab$^{-1}$ and 100 TeV collider with 30 ab$^{-1}$. As seen in Fig.~\ref{fig:Bln}, ML can improve significantly the prospects of heavy-light neutrino mixing angles $|V_{eN}|^2$ and $|V_{\mu N}|^2$ for heavy neutrino mass $m_N \lesssim 1$ TeV in the process in Eq.~(\ref{eqn:trilepton}), by up to roughly one order of magnitude. 
In particular, for $m_N = 100$ GeV, at the HL-LHC 14 TeV with a luminosity of 3 ab$^{-1}$,  the heavy-light neutrino mixing angle $|V_{eN}|^2$ and  $|V_{\mu N}|^2$ can be probed down to $(2.01_{-0.24}^{+0.16}) \times10^{-5}$ and $(1.53_{-0.18}^{+0.18}) \times10^{-5}$, respectively. At the HE-LHC 27 TeV with 15 ab$^{-1}$, the sensitivities of $|V_{eN}|^2$ and  $|V_{\mu N}|^2$ can be improved up to $(6.48_{-0.62}^{+0.83}) \times10^{-6}$ and $(5.28_{-0.48}^{+0.64}) \times10^{-6}$. At the future 100 TeV collider, the prospects are $(3.96_{-0.30}^{+2.18}) \times10^{-6}$  for $|V_{eN}|^2$ and  $(2.75_{-0.54}^{+0.85}) \times10^{-6}$ for $|V_{\mu N}|^2$, respectively. For the large mass $m_N = 1$ TeV, the future 100 TeV collider can even probe the mixing angles $|V_{eN}|^2$ and $|V_{\mu N}|^2$ down to $(2.58_{-0.33}^{+0.61}) \times10^{-4}$ and $(2.68_{-0.42}^{+0.36}) \times10^{-4}$, well below the current EWPD limits. 




\section{Conclusion}
\label{sec:conclusion}

Emerging from many different variations of the seesaw framework, the heavy neutrinos have been one of the well-motivated BSM particles at the high-energy colliders. Depending on the seesaw model details, the heavy neutrinos can be either Dirac or Majorana fermions, or a mixture of them. In this paper, we have studied the application of machine learning to trilepton signals of Dirac heavy neutrino at the high-energy hadron colliders, which can arise for instance in the inverse seesaw model. To be specific, we focus on the charged-current DY process in Eq.~(\ref{eqn:trilepton}), which is mediated by the SM $W$ boson and heavy-light neutrino mixing angles. This process is the dominant production channel at the hadron colliders for heavy neutrino mass $m_N \lesssim 1$ TeV.

With up to one QCD jet, all the kinematic distributions of the charged leptons, missing energy and jets can be found in Figs.~\ref{fig:var_global} to \ref{fig:var_jet}. It turns out that the observables $m_{\ell_1 \ell_2}$ and $m_N^{\rm rc}$ are the most important distinguishing parameters to suppress the SM backgrounds (cf. Fig.~\ref{fig:150_500_ZN}). Other observables, like the reconstructed $W$ boson etc., can also improve the sensitivities especially for $m_N \sim 150$ GeV, where the distributions of observables of the signal events overlap largely with those of the SM events (cf. Figs.~\ref{fig:150_500} and \ref{fig:150_500_ZN}). The distributions of one-dimensional likelihood variables for the SM backgrounds and signal are exemplified in Fig.~\ref{fig:estimator}, and the resultant sensitivities of $m_N$ and $|V_{\ell N}|^2$ (with $\ell = e,\,\mu$) at the 95\% C.L. using machine learning are shown in Fig.~\ref{fig:Bln}, assuming the heavy neutrino mixes with either $\nu_e$ or $\nu_\mu$. For $m_N = 100$ GeV, the mixing angle $|V_{eN}|^2$ ($|V_{\mu N}|^2$) can be probed down to $2.01\times10^{-5}$ ($1.53\times10^{-5}$) at the HL-LHC, $6.48\times10^{-6}$ ($5.28\times10^{-6}$) at the HE-LHC, and $3.96\times10^{-6}$ ($2.75\times10^{-6}$) at the 100 TeV collider, respectively. For $m_N = 1$ TeV, the mixing angles $|V_{eN}|^2$ and $|V_{\mu N}|^2$ can be probed, respectively, down to $2.58 \times10^{-4}$ and $2.68 \times10^{-4}$ at the future 100 TeV collider. This improves significantly the prospects of heavy neutrinos in the trilepton channel at the high-energy hadron colliders.

This paper has been focused only on the case of heavy Dirac neutrinos in the specific channel (\ref{eqn:trilepton}). Machine learning can also be applied to  other production channels such as the gluon fusion and VBF processes, which are important for the heavy neutrino mass above roughly 1 TeV. 
For Majorana heavy neutrinos, there will be LNV same-sign dileptons in the final state. For such signals, the SM backgrounds will be different from those in this paper, and machine learning may also help to improve the prospects of LNV signals at future high-energy colliders. These cases will be pursued in future publications.





\section*{Acknowledgements}

J.F.,\ Y.P.Z.\ and H.H.Z.\ are supported
in part by the National Natural Science Foundation of
China under Grant No.\ 11875327, the Fundamental
Research Funds for the Central Universities, and the Sun
Yat-Sen University Science Foundation. Q.S.Y. is supported by the National Natural Science Foundation of
China under Grant No.\ 11875260. The work of Y.Z.\ is supported by the National Natural Science Foundation of China under Grant No.\  12175039, the 2021 Jiangsu Shuangchuang (Mass Innovation and Entrepreneurship)
Talent Program No.\ JSSCBS20210144, and the ``Fundamental Research Funds for the Central Universities''.

\appendix 
\section{Two solutions for the longitudinal momentum of neutrino}

Considering the process $W^{+} \xrightarrow{} \ell^{+} \nu$ at the hadron colliders, we define the momenta of $\ell^{+}$ and $\nu$ as $p_{\ell}$ and $p_\nu$, respectively. As there might also be additional jets in our analysis, we use $p_{\rm vis}$ to denote the sum of momenta of all the visible particles, including charged leptons and jets in our case, i.e.
\begin{eqnarray}
	p_{\ell} & \ = \ &(E_1,p_{x_1},p_{y_1},p_{z_1}),\\
	p_{\rm vis} & \ = \ &(E_2,p_{x_2},p_{y_2},p_{z_2}),\\
	p_{\nu} & \ = \ &(\sqrt{p_{x_2}^2+p_{y_2}^2+(p_{z}^{\nu})^2}, -p_{x_2}, -p_{y_2},p_{z}^{\nu} ).
\end{eqnarray}
For on-shell $W$ boson decay, the $W$ boson mass $m_W$ satisfies $m_{W}^{2} = (p_\ell+p_\nu)^{2}$, 
from which we obtain a quadratic equation for the longitudinal momentum $p_{z}^{\nu}$ of neutrino:
\begin{eqnarray}
\label{eqn:equation}
a(p_{z}^{\nu})^2 + bp_{z}^{\nu} + c = 0,
\end{eqnarray}
where,
\begin{eqnarray}
a& \ = \ &4p_{z_1}^2-4E_1^2 \,,\\
b& \ = \ &4c_0p_{z_1} \,,\\
c& \ = \ &c_0^2-4E_1^2(p_{x_2}^2+p_{y_2}^2) \,,
\end{eqnarray}
with 
\begin{eqnarray}
c_0 \ = \ m_{W}^2-p_{\ell}^2-2p_{x_1}p_{x_2}-2p_{y_1}p_{y_2} \,.
\end{eqnarray}
When $b^2-4ac \geq 0$ the equation (\ref{eqn:equation}) has two real solutions. The solution with the larger absolute value is labelled as $p_z^{\nu,0}$ and the other one as $p_z^{\nu,1}$. If $b^2-4ac < 0$, the two solutions will be complex. Then we will keep only the real parts. 

\bibliographystyle{JHEP}
\bibliography{ref}

\end{document}